\documentclass[a4paper,fleqn,usenatbib]{mnras}
\hypersetup{draft}

\usepackage{newtxtext,newtxmath}

\usepackage[T1]{fontenc}
\usepackage{ae,aecompl}

\usepackage{graphicx}	
\usepackage{amsmath}	
\usepackage{amssymb}	
\usepackage{lineno,hyperref}
\usepackage{pdflscape}
\usepackage{multirow}

\usepackage{siunitx}
\usepackage[T1]{fontenc}
\usepackage[utf8]{inputenc}
\usepackage{gensymb}
\usepackage{wasysym}

\usepackage{esvect}

\newcommand{\percent}{\% }

\usepackage[colorinlistoftodos]{todonotes}
\let\Oldtodo\todo
\renewcommand{\todo}[1]{\Oldtodo[inline]{#1}}

\title[Novel meteor trajectory solver]{Estimating trajectories of meteors: an observational Monte Carlo approach - II. Results}

\author[D. Vida et al.]{
Denis Vida,$^{1,2}$\thanks{E-mail: dvida@uwo.ca}
Peter G. Brown,$^{2,3}$
Margaret Campbell-Brown,$^{2,3}$
Paul Wiegert,$^{2,3}$
\newauthor
Peter S. Gural$^{4}$
\\
$^{1}$Department of Earth Sciences, University of Western Ontario, London, Ontario, N6A 5B7, Canada\\
$^{2}$Department of Physics and Astronomy, University of Western Ontario, London, Ontario, N6A 3K7, Canada\\
$^{3}$Centre for Planetary Science and Exploration, University of Western Ontario, London, Ontario, N6A 5B8, Canada\\
$^{4}$Gural Software and Analysis LLC, Sterling, Virginia, 20164 USA\\
}

\date{Accepted 2019 November 25. Received 2019 November 10; in original form 2019 September 23.}

\pubyear{2019}

\begin{document}
\label{firstpage}
\pagerange{\pageref{firstpage}--\pageref{lastpage}}
\maketitle

\begin{abstract}

In the first paper of this series we examined existing methods of optical meteor trajectory estimation and developed a novel method which simultaneously uses both the geometry and the dynamics of meteors to constrain their trajectories. We also developed a simulator which uses an ablation model to generate realistic synthetic meteor trajectories which we use to test meteor trajectory solvers. In this second paper, we perform simulation validation to estimate radiant and velocity accuracy which may be achieved by various meteor observation systems as applied to several  meteor showers.

For low-resolution all-sky systems, where the meteor deceleration is generally not measurable, the multi-parameter fit method assuming a constant velocity better reproduces the radiant and speed of synthetic meteors. For moderate field of view systems, our novel method performs the best at all convergence angles, while multi-parameter fit methods generally produce larger speed errors. For high-resolution, narrow field of view systems, we find our new method of trajectory estimation reproduces radiant and speed more accurately than all other methods tested. The ablation properties of meteoroids are commonly found to be the limiting factor in velocity accuracy.

We show that the true radiant dispersion of meteor showers can be reliably measured with moderate field of view (or more precise) systems provided appropriate methods of meteor trajectory estimation are employed. Finally, we compare estimated and real angular radiant uncertainty and show that  for the solvers tested the real radiant error is on average underestimated by a factor of two.

\end{abstract}




\section{Introduction} \label{sec:introduction}

This paper is a direct continuation of an earlier work \citep[][hereafter Paper 1]{vida2019meteortheory}, in which we developed both a new method for estimating meteor trajectories and a meteor trajectory simulator. Paper 1 also presented a summary of the theory behind earlier meteor trajectory determination algorithms. In this paper we attempt to answer the following question: For a given type of optical meteor observation system, what is the best trajectory solver to use, and what is the associated expected quantitative accuracy? We note that this is only the first step in the process of estimating a meteoroid's original heliocentric orbit. The necessary additional step is accounting for deceleration due to atmospheric drag prior to the earliest measured luminous point of the meteor, a topic addressed in \cite{vida2018modeling}.

In Paper 1 we analyzed the shortcomings of existing methods of meteor trajectory estimation, with particular focus on application to the high-precision data collected by the Canadian Automated Meteor Observatory's (CAMO) mirror tracking system \citep{weryk2013camo}. In Paper 1 we examined the most commonly used meteor trajectory estimation methods in detail, including: the geometrical intersecting planes \citep{ceplecha1987geometric} and lines of sight \citep{borovicka1990comparison} approaches, and the multi-parameter fit (MPF) method by \cite{gural2012solver}. As pointed out by \cite{egal2017challenge}, the true measurement accuracy of these methods for various meteor observation systems and showers is unknown, and the most advanced of them, the MPF method with the exponential deceleration velocity model, is numerically problematic to fit to observations.

In an attempt to improve on existing algorithms, we developed a novel meteor trajectory estimation method which uses both the geometry and the dynamics of a meteor to constrain a trajectory solution, but without an assumed underlying  kinematic model. We also developed a meteor trajectory simulator which uses the meteor ablation model of \cite{campbell2004model} to simulate realistic dynamics and light curves of meteors given their physical properties as a means to compare and test meteor trajectory solvers.

In this work, we apply the simulator and explore the accuracy of each trajectory solver to three types of typical optical meteor observation systems: a low-resolution all-sky system, a moderate field of view system, and the high-precision CAMO mirror tracking system. For each system we used the simulator to investigate the ability of each solver to properly recover the geocentric radiant and velocity of three major showers spanning a wide range of meteor velocities and meteoroid types (Draconids, Geminids, Perseids). The parameters used for simulations and the comparison between simulations and real-world observations are given in section \ref{sec:modelling_details}. We also perform dynamical modelling of the 2011 Draconid outburst, which  was produced by recently ejected meteoroids  \citep{vaubaillon2011coming} and thus should have a very tight radiant. We use this compact shower to estimate the radiant measurement accuracy needed to resolve the true physical dispersion of a meteor shower.

In section \ref{sec:results} we present simulation results and compare the performance of various meteor trajectory estimation methods across all simulated meteor observation systems and our three chosen showers. In section \ref{subsec:2015taurids} we examine solver performances as applied to a specific case study, namely the unique 2015 Taurid outburst. This outburst was arguably the first instance  where we have both strong a priori knowledge of the expected orbits (particularly semi-major axis) and a large number of high - precision meteor trajectories \citep{spurny2017discovery}. We also consider the special case of long duration fireballs where the influence of gravity is particularly important by simulating solver performance for an all-sky system as will be discussed in section \ref{subsec:long_fireballs}. Finally, in section \ref{subsec:error_analysis} we examine the accuracy of meteor trajectory error estimation by comparing estimated radiant errors to offsets from the simulated ground truth.

\section{Simulation-based performance analysis of trajectory solvers} \label{sec:modelling_details}

\subsection{Hardware models}

To compare the performance of various existing meteor trajectory solvers with the new Monte Carlo method, we appeal to simulations. The method of generating simulated meteor observations is described in detail in Paper 1 \citep{vida2019meteortheory}. We simulated three optical meteor observation systems to generate synthetic meteors to feed into each trajectory simulator. These three systems follow the optical model system choices previously discussed in \cite{vida2018modeling}. The characteristics of these systems (which largely vary in terms of angular precision) include:

\begin{enumerate}

  \item A low resolution all-sky CCD video fireball system based on the hardware of the Southern Ontario Meteor Network (SOMN) \citep{brown2010development}.
  \item A moderate field of view CCD video system typical of CAMS \citep{jenniskens2011cams}, SonotaCo\footnote{SonotaCo: \url{http://sonotaco.jp/}}, the Croatian Meteor Network \citep{gural2009new}, and the Global Meteor Network \citep{vida2019overview}.
  \item An image intensified mirror tracking system based on the Canadian Automated Meteor Observatory (CAMO) \citep{weryk2013camo}.

\end{enumerate}
  
\noindent These systems cover a wide range of observed meteoroid masses, fields of view, and astrometric precision. Details of each system are given in table \ref{tab:system_parameters}.

Our simulated all-sky fireball network consisted of 3 stations in an equilateral triangle configuration with \SI{100}{\kilo \metre} long sides (stations A1, A2, A3 in the simulation). The cameras at each station are pointing straight up and have a field of view (FOV) of $\ang{120} \times \ang{120}$. Larger FOVs were difficult to simulate as the volume of the sky that needed to be randomly sampled becomes very high, and most of it was outside the FOV of other cameras. The measurement uncertainly was assumed to be 2 arc minutes and the frames per second (FPS) of the cameras 30. 

For the CAMS-like moderate FOV system, we also chose to use 3 stations in the equilateral triangle configuration (stations M1, M2, M3 in the simulation). These had FOVs of $\ang{64} \times \ang{48}$, 30 FPS and a measurement uncertainly of 30 arc seconds. The elevation of the centres of the fields of view of all cameras was \ang{65} and they where all pointed towards the centre of the triangle. 

Finally, the simulated CAMO system mimics the real system which has 2 stations (``tavis'' and ``elgin'' in the simulation) separated by \SI{45}{\kilo \metre}, a FOV of $\ang{30} \times \ang{30}$, cameras operated at 100 FPS, and a precision of 1 arc second.

\begin{table*}
    \caption{Parameters of simulated optical meteor observation systems. $\Delta t_{max}$ is the maximum time offset, FPS is the frames per second of the camera, $\sigma_{obs}$ the measurement uncertainly in arc seconds, FOV width and height are the size of the field of view in degrees, MLM the meteor limiting magnitude, and $P_{0m}$ is the power of a zero-magnitude meteor \protect\citep[power values taken from][]{weryk2013simultaneous}.}
    {
    \begin{tabular}{l c c c c c c c c c c c}
    \hline\hline 
    System & N stations & $\Delta t_{max} (\SI{}{\second}$) & FPS & $\sigma_{obs}$ (arcsec) & FOV width (deg) & FOV height (deg) & MLM & $P_{0m}$ (\SI{}{\watt})\\
    \hline 
    CAMO     & 2 & 1 & 100 &   1 &  30 &  30 &  +5.5 & 840 \\
    Moderate & 3 & 1 & 30  &  30 &  64 &  48 &  +5.0 & 1210 \\
    All-sky   & 3 & 1 & 30  & 120 & 120 & 120 & -0.5 & 1210 \\
    \hline 
    \end{tabular}
    }
    \label{tab:system_parameters}
\end{table*}

\subsection{Simulated meteor showers} \label{subsec:simulated_showers}

To explore the performance of various meteor trajectory estimation methods when observing meteors of different velocities and physical properties we focused on generating synthetic meteors from three very different meteor showers. We simulated 100 meteors for every system for each of the following three meteor showers: 

\begin{enumerate}
    \item{The 2011 Draconids, a low-velocity ($\SI{\sim 21}{\kilo \metre \per \second}$) shower with fragile and fresh (<100 years of age) cometary meteoroids \citep{borovivcka2007atmospheric} that experienced an outburst in 2011 \citep{segon2014draconids, ye2013radar}}.
    \item{The 2012 Geminids, a $\SI{\sim 34}{\kilo \metre \per \second}$ moderate speed shower of asteroidal origin containing meteoroids of ages from 1000 to 4000 years \citep{beech2002age}.}
    \item{The 2012 Perseids, a $\SI{\sim 59}{\kilo \metre \per \second}$ fast shower of Halley-type comet origin whose meteoroids were ejected  >2000 years ago \citep{brown1998simulation}.}
\end{enumerate}

\noindent Realistic trajectories and dynamics were simulated using the \cite{campbell2004model} meteoroid ablation model procedure described in detail in \cite{vida2018modeling}. Meteor shower parameters used in the simulations are given in table \ref{tab:shower_parameters} - parameters of all showers except the Draconids were taken from observations published in the literature.

Note that the 2015 Taurid fireball outburst was also simulated, but only for the all-sky systems as discussed in section \ref{subsec:2015taurids}. The goal in applying our analysis to the unique 2015 Taurid outburst  was to contrast the accuracy of various trajectory estimation methods when using low-precision (video) all-sky systems as compared to higher precision fireball systems.

\subsection{Dynamical modelling of the 2011 Draconid outburst}

The 2011 Draconids were the youngest of the simulated showers and should have the most compact radiant. The measured radiant spread should be dominated by measurement uncertainty when measured with less precise systems. To quantify the minimum accuracy required to observe the true physical radiant and velocity dispersion of the 2011 Draconids, we appeal to dynamical modelling of the shower. Here we use the method of \cite{wiegert2009dynamical} to obtain an estimate of both the true average location of the radiant and velocity of the outburst and its theoretical spread. We then use these as inputs to our simulation model to generate synthetic 2011 Draconids to virtually "observe" with each of our three optical systems and apply each meteor trajectory solver in turn.

To dynamically model the 2011 Draconid outburst, the orbital elements of the 1966 apparition of 21P/Giacobini-Zinner were integrated backwards 200 years with the RADAU \citep{everhart85RADAU} integrator within a simulated Solar System containing the Sun and eight planets. The parent comet was then advanced forward in time while ejecting meteoroids with radii between \SI{100}{\micro \metre} and \SI{10}{\centi \metre} when within 3 AU of the Sun. The ejection speed and direction follows the approach of the \cite{brown1998simulation} model with an assumed comet radius of \SI{1}{\kilo \metre}, albedo of 0.05 and bulk density of \SI{300}{\kilo \gram \per \cubic \metre}. 

Meteoroids arriving at Earth in 2011 were found to be produced by the 1838 and 1907 comet perihelion passages, with smaller contributions from 1920 and 1953. The simulated peak coincided with that reported by visual observers to the International Meteor Organization (IMO) Visual Meteor Database\footnote{IMO VMDB 2011 Draconids: \url{https://www.imo.net/members/imo_live_shower?shower=DRA&year=2011}}. 

\begin{figure}
  \includegraphics[width=\linewidth]{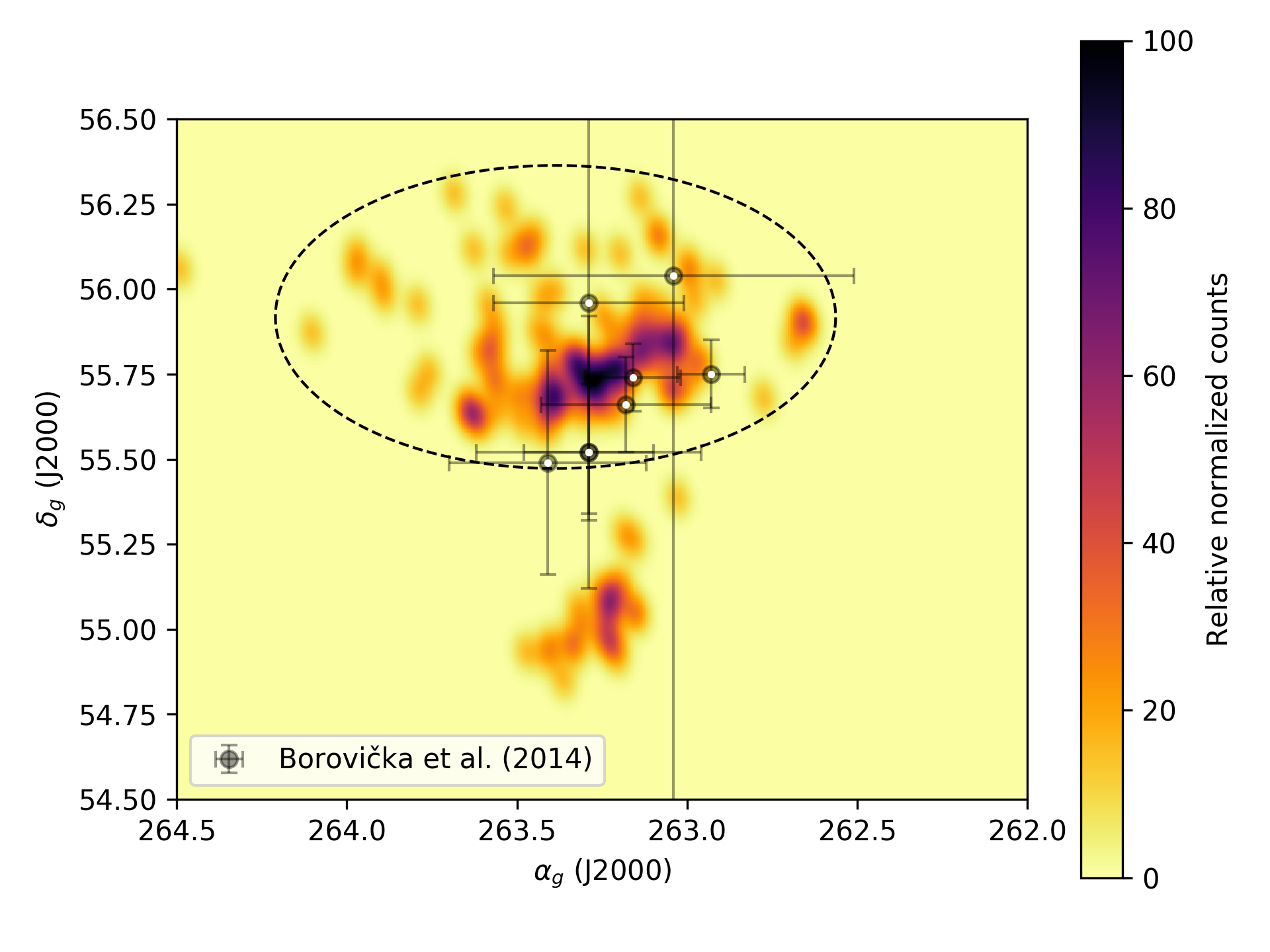}
  \caption{Density map of simulated geocentric equatorial (J2000.0) radiants of the 2011 Draconids at the time of peak activity. A bi-variate Gaussian was fit to the radiants ($\alpha_g = \ang{263.387} \pm \ang{0.291}, \delta_g = \ang{55.9181} \pm \ang{0.158}$). The corresponding $2\sigma$ level is shown as a black contour. Draconid radiants observed by \protect\cite{borovivcka2014spectral} in 2011 are also shown.}
  \label{fig:2011draconids_sim}
\end{figure}

The radiants of the dynamically modelled stream that impacted the Earth are shown in figure \ref{fig:2011draconids_sim}. Note that these model radiants are without observational biases because they were directly computed from simulated meteoroids arriving at Earth. The position and the dispersion of the modelled radiant at the time of peak activity was $\alpha_g = \ang{263.387} \pm \ang{0.291}, \delta_g = \ang{55.9181} \pm \ang{0.158}$. The values were derived by fitting a bi-variate Gaussian to the modelled radiants. Distinct radiant structure can also be seen - we estimate that an observational radiant precision of better than 3 to 6 arc minutes (\ang{0.05} - \ang{0.1}) is needed for the true radiant structure to be unambiguously reconstructed from observations. We use these values as the absolute minimum radiant accuracy needed to resolve the true physical radiant spread for showers with the most compact radiants.

The video and high power large aperture (HPLA) radar observations of the Draconid outburst measured an almost order of magnitude larger dispersion than our model predicts, suggesting they did not record the intrinsic (physical) radiant spread of the shower \citep{segon2014draconids, trigo20132011, kero2012mu}. We note that the video observations of the outburst incorporating high-quality manual reductions reported by \cite{borovivcka2014spectral} are an excellent match to our simulations in both radiant position, dispersion, and the simulated peak time, as shown in figure \ref{fig:2011draconids_sim}.

As both the Geminids and Perseids are older showers, we do not expect the physical radiant dispersion to be as compact as the Draconids, thus for simulation purposes we have used observed values for these quantities provided in \cite{jenniskens2016established} and \cite{jenniskens1994meteor}. We summarize the modeling parameters adopted of the simulated meteors showers in table \ref{tab:shower_parameters} and the physical properties of shower meteoroids used in the ablation modeling in table \ref{tab:meteoroid_properties}.

\begin{table*}
    \caption{The model parameters adopted for simulated meteor showers. The parameters for the Draconids were computed from the simulation as discussed in the text. The parameters for the Geminids and Perseids were taken from \protect\cite{jenniskens2016established} unless otherwise noted. Parameters for the 2015 Taurids were taken from \protect\cite{spurny2017discovery}. These  were modified from the original so that they are centered around the peak solar longitude while the radiant spread was computed directly from data provided in \protect\cite{spurny2017discovery}. $\lambda_{\astrosun}^{max}$ is the solar longitude of the peak (degrees), $B$ is the solar longitude slope of the rising portion of the activity profile following the procedure of \protect\cite{jenniskens1994meteor}, $\alpha$ is the mean geocentric right ascension, $\Delta \alpha$ is the radiant drift (degree on the sky per degree of solar longitude), $\alpha_\sigma$ is the standard deviation in R.A., $\delta$ is the mean geocentric declination, $\Delta \delta$ is the declination radiant drift, $\delta_\sigma$ is the standard deviation of the declination, $V_g$ is the mean geocentric velocity in \SI{}{\kilo \metre \per \second}, $\Delta V_g$ is the change in geocentric velocity per degree of solar longitude, and $V_{g\sigma}$ is the standard deviation of the geocentric velocity.}
    {
    \begin{tabular}{l c c c c c c c c c c c c}
    \hline\hline 
    Shower    & Year & $\lambda_{\astrosun}^{max}$ & $B$      & $\alpha$  & $\Delta \alpha$ & $\alpha_\sigma$ & $\delta$   & $\Delta \delta$ & $\delta_\sigma$ & $V_g$   & $\Delta V_g$ & $V_{g\sigma}$ \\
    \hline 
    Draconids & 2011 & 198.07 & 17.5, 1*        & 263.39 & 0.0   & 0.29 & 55.92 & 0.0   & 0.16 & 20.93 & 0     & 0.04 \\
    Geminids  & 2012 & 262.0  & $\sim 0.5$, 2*  & 113.5  & 1.15  & 2.8  & 32.3  & -0.16 & 1.5  & 33.8  & 0     & 2.0 \\
    Perseids  & 2012 & 140.0  & 0.4, 2*         & 48.2   & 1.4   & 2.8  & 58.1  & 0.26  & 1.7  & 59.1  & 0     & 2.4 \\
    Taurids - resonant branch   & 2015 & 221.0  & 0.15            & 53.06  & 0.554 & 0.33 & 14.66 & 0.06  & 0.27 & 29.69 & -0.293 & 0.22 \\
    \hline 
    \end{tabular}
    }\\
    1* - \citep{koten2014three}, 2* - \citep{jenniskens1994meteor}
    \label{tab:shower_parameters}
\end{table*}

\begin{table*}
	\caption{Physical properties of meteoroids adopted as input to the ablation model in simulating our four meteor showers. Here $s$ is the mass index, $\rho$ is the range of bulk densities of meteoroids, $\sigma$ is the apparent ablation coefficient \citep{ceplecha2005}, and $L$ is the energy needed to ablate a unit of meteoroid mass.}
    {
	\begin{tabular}{l c c c c}
	\hline\hline 
	Shower & $s$ & $\rho$ (\SI{}{\kilogram \per \cubic \metre}) & $\sigma$ (\SI{}{\square \second \per \square \kilo \metre}) & $L$ (\SI{}{\joule \per \kilogram})\\
	\hline 
	Draconids  & 1.95, 1* & 100 - 400, 2*        & 0.21, D-type, 3*  & \num{1.2d6} \\
	Geminids   & 1.7, 4*  & 1000 - 3000, 5*      & 0.042, A-type, 3* & \num{6.0d6} \\
	Perseids   & 2.0,  7* & HTC distribution, 6* & 0.1, C-type, 3*   & \num{2.5d6} \\
	Taurids    & 1.8,  8* & 1200 - 1600, 9*      & 0.1, C-type, 3*   & \num{2.5d6} \\
	\hline 
	\end{tabular}
	}
	\\1* - \citep{koten2014three}, 2* - \citep{borovivcka2007atmospheric}, 3* - \citep{ceplecha1998meteor}, 4* - \citep{blaauw2011meteoroid}, 5* - \citep{borovivcka2009material}, 6* - \citep{moorhead2017two}, 7* - \citep{beech1999search}, 8* - \citep{moser2011luminous}, 9* - \citep{brown2013meteorites}
	\label{tab:meteoroid_properties}
\end{table*}

\subsection{Simulation validation} \label{subsec:sim_obs_comparison}

To confirm the appropriateness of the simulations, we compare some metrics among our suite of simulated and observed meteors for the same optical system. Note that we did not attempt to reconstruct particular observed events though simulation; we only identified meteors of similar properties and quantitatively compared the trajectory fit residuals and deceleration. As an indicator of deceleration we computed the meteor's lag, i.e. how much the observed meteor falls behind a hypothetical meteoroid moving with a fixed speed equal to the initial velocity. We present several meteors from instrument data sets having comparable speed and duration to those we simulated. Table \ref{tab:comparison_meteors_params} compares the initial speed, mass and zenith angle of a selection of simulated and representative observed meteors. All observations were reduced using the Monte Carlo method.

\begin{figure*}
  \includegraphics[width=\linewidth]{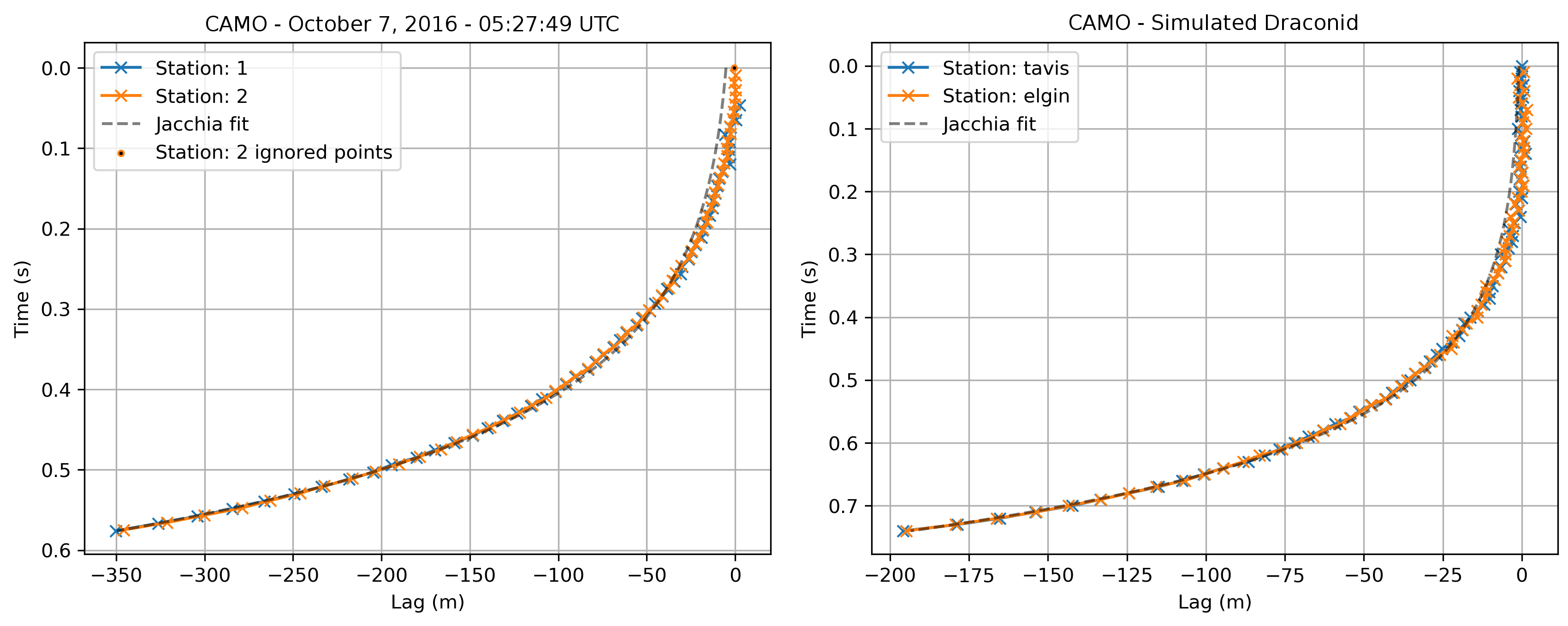}
  \caption{Left: Lag of a CAMO sporadic meteor observed on October 7, 2016 with $V_g$ = \SI{21.4}{\kilo \metre \per \second}. As per the UWO station naming convention, the stations identifiers are numbers - 1 is Tavistock, 2 is Elginfield. Right: A simulated CAMO Draconid of similar mass and with $V_g$ = \SI{20.9}{\kilo \metre \per \second}.}
  \label{fig:camo_obs_sim_comp}
\end{figure*}

\begin{figure*}
  \includegraphics[width=\linewidth]{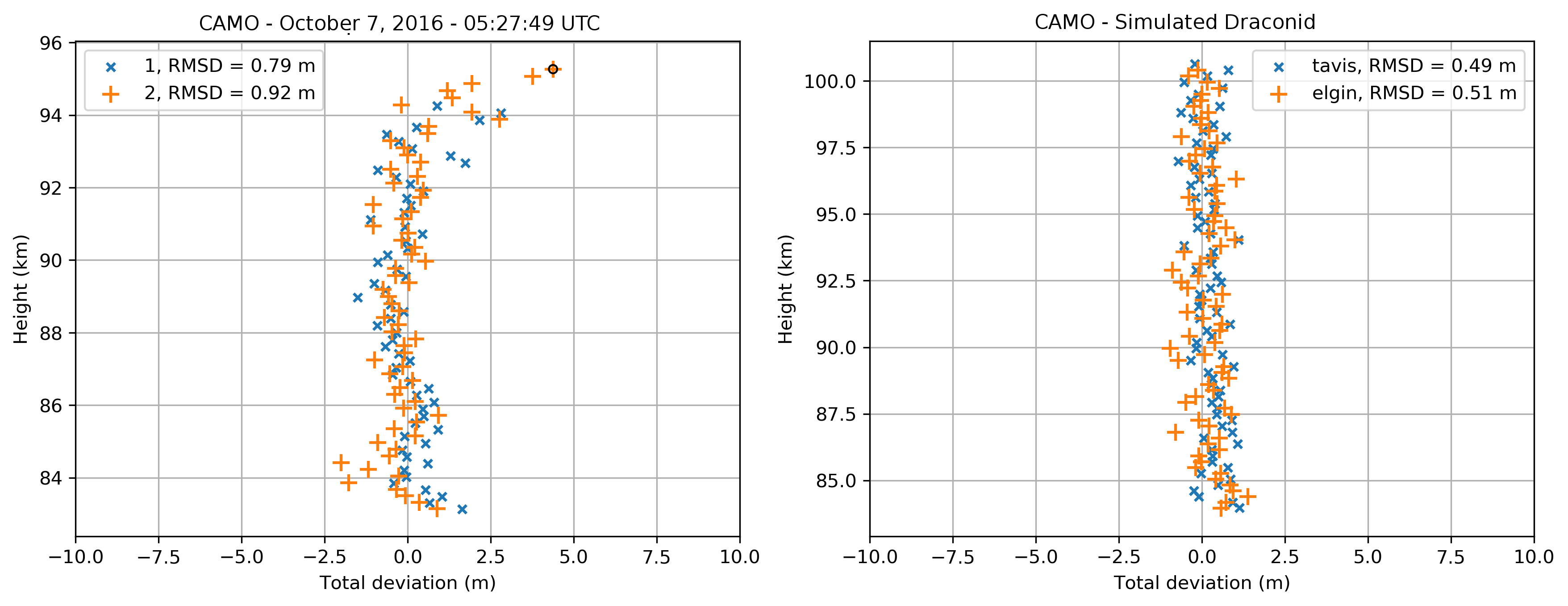}
  \caption{Left: Spatial residuals of a CAMO sporadic meteor observed on October 7, 2016 with $V_g$ = \SI{21.4}{\kilo \metre \per \second}. Station 1 is Tavistock, 2 is Elginfield. Right: A simulated CAMO Draconid of similar mass and with $V_g$ = \SI{20.9}{\kilo \metre \per \second}.}
  \label{fig:camo_obs_sim_comp_residuals}
\end{figure*}

As a first example, figures \ref{fig:camo_obs_sim_comp} and \ref{fig:camo_obs_sim_comp_residuals} show a sporadic meteor with a geocentric velocity $V_g = \SI{\sim 21}{\kilo \metre \per \second}$ observed with CAMO and comparable simulated CAMO Draconid. The "Jacchia fit" curve in lag plots is a fit of the exponential deceleration model of \cite{whipple1957reduction} to the computed lag and is only used for visualization purposes. The amount of deceleration and the scatter in the spatial fit residuals (\SI{< 1}{\metre}) are similar. The scatter in residuals shows that the dispersion due to random errors are comparable. CAMO measurements are slightly noisier compared to the model at the beginning because the tracking mirrors require several milliseconds to settle. The fit residuals were on the order of 1 arc second for both data sets. We note that the observed meteor showed significant visible fragmentation which was not included in the model; thus the magnitude of the observed lag is larger than in the simulation.

\begin{figure*}
  \includegraphics[width=\linewidth]{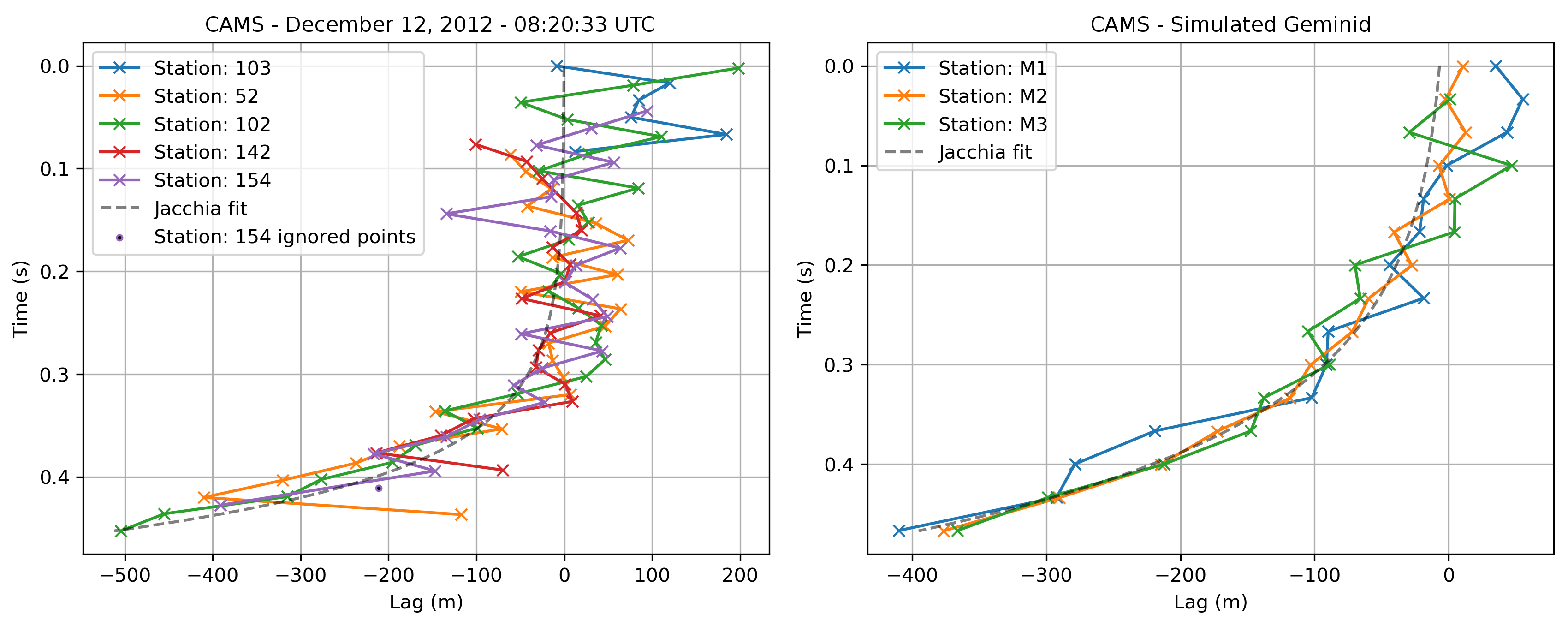}
  \caption{Left: Lag of a CAMS Geminid observed on December 12, 2012 with $V_g$ = \SI{33.4}{\kilo \metre \per \second} by CAMS cameras in California. Right: A simulated CAMS Geminid of similar mass with $V_g$ = \SI{34.6}{\kilo \metre \per \second}.}
  \label{fig:cams_obs_sim_comp}
\end{figure*}

Figure \ref{fig:cams_obs_sim_comp} shows the lag for a Geminid observed by five CAMS stations in California compared to a simulated CAMS Geminid. The two meteors had similar velocities, masses, and entry angles. Both meteors show similar decelerations, and had spatial fit residuals of $\SI{\sim 20}{\metre}$.

\begin{figure*}
  \includegraphics[width=\linewidth]{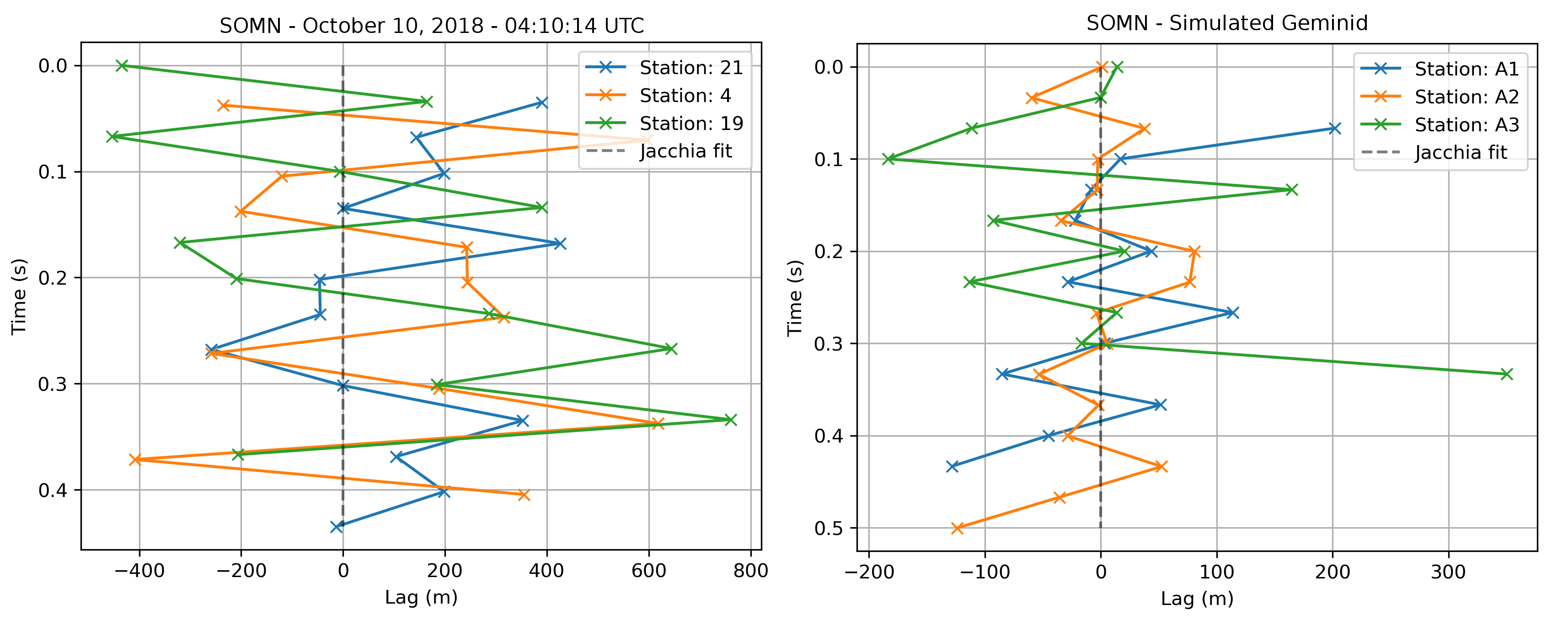}
  \caption{Left: Lag of a SOMN Southern Taurid meteor observed on October 10, 2018 with $V_g$ = \SI{31.4}{\kilo \metre \per \second}. Ignored points are those with angular error more than $3\sigma$ above the mean after the first trajectory estimation pass as described in Paper 1. Right: A simulated SOMN Geminid with $V_g$ = \SI{34.0}{\kilo \metre \per \second}.}
  \label{fig:somn_obs_sim_comp}
\end{figure*}

Finally, figure \ref{fig:somn_obs_sim_comp} shows the comparison for a low-resolution, all-sky system, namely the Southern Ontario Meteor Network \cite{brown2010development} between an observed Southern Taurid ($V_g$ = \SI{31.4}{\kilo \metre \per \second}) and a simulated Geminid. Neither meteor shows visible deceleration due to the low precision of the measurements. However, deceleration may become visible and significant for much longer duration fireballs (see section \ref{subsec:long_fireballs}).

\begin{table*}
    \caption{Comparison of several selected observed meteors and the close fits from among the simulated set of meteors. The photometric masses were computed using a dimensionless luminous efficiency of $\tau = 0.7 \percent$ \protect\citep{vida2018modeling} and a bandpass specific $P_{0m} = \SI{1210}{\watt}$. The observed (calculated) masses are similar to simulated masses. Note that the range of simulated masses was taken from \protect\cite{vida2018modeling} and is based on the masses calculated from observations by each type of system.}
    {
    \begin{tabular}{l c c c c c c}
    \hline\hline 
    \multicolumn{1}{c}{\multirow{2}{*}{System}} & \multicolumn{2}{c}{Initial speed (km/s)} & \multicolumn{2}{c}{Mass (g)} & \multicolumn{2}{c}{Zenith angle (deg)} \\
    \multicolumn{1}{c}{} & Obs & Sim & Obs & Sim & Obs & Sim \\ 
    \hline 
    CAMO     & 24.1 & 23.7 & 0.03 & 0.03 & 26 & 15 \\
    Moderate & 35.2 & 36.4 & 0.02 & 0.01 & 22 & 21 \\
    All-sky   & 33.3 & 36.4 & 0.2  & 0.1  & 40 & 36 \\
    \hline 
    \end{tabular}
    }
    \label{tab:comparison_meteors_params}
\end{table*}

\section{Results} \label{sec:results}

Following the theoretical development given in Paper 1, we numerically evaluated the performance of the following trajectory solvers (abbreviations used later in the text):
\begin{itemize}
    \item IP - The intersecting planes method of \cite{ceplecha1987geometric}. The initial velocity is computed as the average velocity of the first half of the trajectory; better initial velocity accuracy might be achieved by using the method of \cite{pecina1983new, pecina1984importance}.
    \item LoS - Our implementation of the \cite{borovicka1990comparison} method with our progressive initial velocity estimation method as described in Paper 1.
    \item LoS-FHAV - Our implementation of the \cite{borovicka1990comparison} method. The initial velocity is computed as the average velocity of the first half of the trajectory.
    \item MC - The Monte Carlo method presented in Paper 1.
    \item MPF const - The multi-parameter fit method of \cite{gural2012solver} using a constant velocity model.
    \item MPF const-FHAV - For this hybrid-solver, the radiant solution is taken from the MPF constant velocity model, but the lines of sight are re-projected on the trajectory and the initial velocity is estimated as the slope of the length vs. time along the track (effectively, the average velocity) of the first half of the trajectory.
    \item MPF linear - The \cite{gural2012solver} multi-parameter fit method with a linear deceleration velocity model.
    \item MPF exp - The \cite{gural2012solver} multi-parameter fit method with the exponential deceleration model of \cite{whipple1957reduction}
\end{itemize}

Global results for all tested trajectory solvers for all-sky systems are given in table \ref{tab:allsky_solvers_performance}, for moderate FOV systems in table \ref{tab:cams_solvers_performance}, and for CAMO in table \ref{tab:camo_solvers_performance}  respectively. For every combination of observation system, meteor shower and trajectory solver we list:

\begin{enumerate}
    \item The column labelled F in each table is the total number of failures for a given method out of 100 simulated runs. This is the number of trajectory solutions with radiant or velocity difference between the true (simulated) and estimated values larger than the predetermined values given in the caption to each table.
    \item The standard deviation between the estimated and true radiant angular separation ($\sigma_R$ in tables).
    \item The standard deviation between the estimated and true geocentric velocity ($\sigma_V$ in tables). 
\end{enumerate}

\noindent The standard deviations are computed after iteratively rejecting solutions outside $3 \sigma$.

A trajectory solution was considered to be a failure if the radiant error (difference between estimated and true as initially input into the simulation) was more than $\Delta_{Rmax}$ degrees from the true radiant, or if the velocity error was more than $\Delta_{Vmax}$ from the model velocity. For the simulated all-sky system the values used were $\Delta_{Rmax} = \ang{5}$, $\Delta_{Vmax} = \SI{5}{\kilo \metre \per \second}$, while for the moderate FOV (CAMS-like) system we adopted $\Delta_{Rmax} = \ang{1}$, $\Delta_{Vmax} = \SI{1}{\kilo \metre \per \second}$. Finally,  for the simulated CAMO--like system, we adopted $\Delta_{Rmax} = \ang{0.5}$, $\Delta_{Vmax} = \SI{0.5}{\kilo \metre \per \second}$. 

Solutions were also removed from further consideration if any of the multi-station convergence angles were less than \ang{15}, \ang{10}, and \ang{1} for the all-sky, CAMS, and CAMO simulations respectively. This procedure was adopted so that the general performance of different solvers can be compared, excluding excessive deviations due simply to low convergence angles. We explore in more depth the dependence of solution accuracy on the convergence angle in section \ref{subsec:convergence_angle_study}.

In what follows, we show representative plots of the spread in the radiant and velocity accuracy for each trajectory solver for each optical system. In addition, we show a selection of individual results per shower and per solver in the form of 2D histograms (e.g. figure \ref{fig:allsky_solver_selection}) which highlight the scatter of estimates among the 100 simulated meteors. On these plots, the angular distance between the real and the estimated geocentric radiant is shown on the X axis, the error in the geocentric velocity is shown on the Y axis, and the bin count is color coded (darker color means higher count).

\subsection{All-sky systems}

\begin{table*}
	\caption{Comparison of solver accuracy for a simulated three station all-sky system. The trajectory was taken to be valid for simulation if the converge angle was larger than \ang{15}. F is the number of failures (out of 100), i.e. the number of radiants that were outside the window bounded by $\Delta_{Rmax} = \ang{5}$, $\Delta_{Vmax} = \SI{5}{\kilo \metre \per \second}$.}
    {
	\begin{tabular}{l | c | c | c | c | c | c | c | c | c}
	\hline\hline 
	Solver           & \multicolumn{3}{|c|}{DRA} & \multicolumn{3}{|c|}{GEM} & \multicolumn{3}{|c}{PER} \\
	                 & F  & $\sigma_R$ & $\sigma_V$ & F & $\sigma_R$ & $\sigma_V$ & F & $\sigma_R$ & $\sigma_V$ \\
	\hline
IP                &  1 & \ang{0.44} & \SI{0.38}{\kilo \metre \per \second} &  1 & \ang{0.21} & \SI{0.38}{\kilo \metre \per \second} & 10 & \ang{0.67} & \SI{0.99}{\kilo \metre \per \second}\\
LoS               &  1 & \ang{0.56} & \SI{0.47}{\kilo \metre \per \second} &  1 & \ang{0.26} & \SI{0.42}{\kilo \metre \per \second} &  3 & \ang{0.66} & \SI{0.46}{\kilo \metre \per \second}\\
LoS-FHAV          &  1 & \ang{0.51} & \SI{0.38}{\kilo \metre \per \second} &  1 & \ang{0.26} & \SI{0.38}{\kilo \metre \per \second} &  7 & \ang{0.66} & \SI{0.53}{\kilo \metre \per \second}\\
Monte Carlo       &  2 & \ang{0.52} & \SI{0.36}{\kilo \metre \per \second} &  1 & \ang{0.24} & \SI{0.47}{\kilo \metre \per \second} &  4 & \ang{0.80} & \SI{0.38}{\kilo \metre \per \second}\\
MPF const         &  1 & \ang{0.28} & \SI{0.46}{\kilo \metre \per \second} &  0 & \ang{0.24} & \SI{0.84}{\kilo \metre \per \second} &  1 & \ang{0.47} & \SI{0.37}{\kilo \metre \per \second}\\
MPF const-FHAV    &  1 & \ang{0.27} & \SI{0.38}{\kilo \metre \per \second} &  0 & \ang{0.23} & \SI{0.41}{\kilo \metre \per \second} &  4 & \ang{0.35} & \SI{0.80}{\kilo \metre \per \second}\\
MPF linear        &  0 & \ang{0.26} & \SI{0.54}{\kilo \metre \per \second} &  4 & \ang{0.23} & \SI{1.76}{\kilo \metre \per \second} &  6 & \ang{0.35} & \SI{0.60}{\kilo \metre \per \second}\\
MPF exp           &  0 & \ang{0.41} & \SI{1.39}{\kilo \metre \per \second} &  1 & \ang{0.21} & \SI{1.65}{\kilo \metre \per \second} & 12 & \ang{0.32} & \SI{1.18}{\kilo \metre \per \second}\\
	\hline 
	\end{tabular}
	}
	\label{tab:allsky_solvers_performance}
\end{table*}

\begin{figure}
  \includegraphics[width=\linewidth]{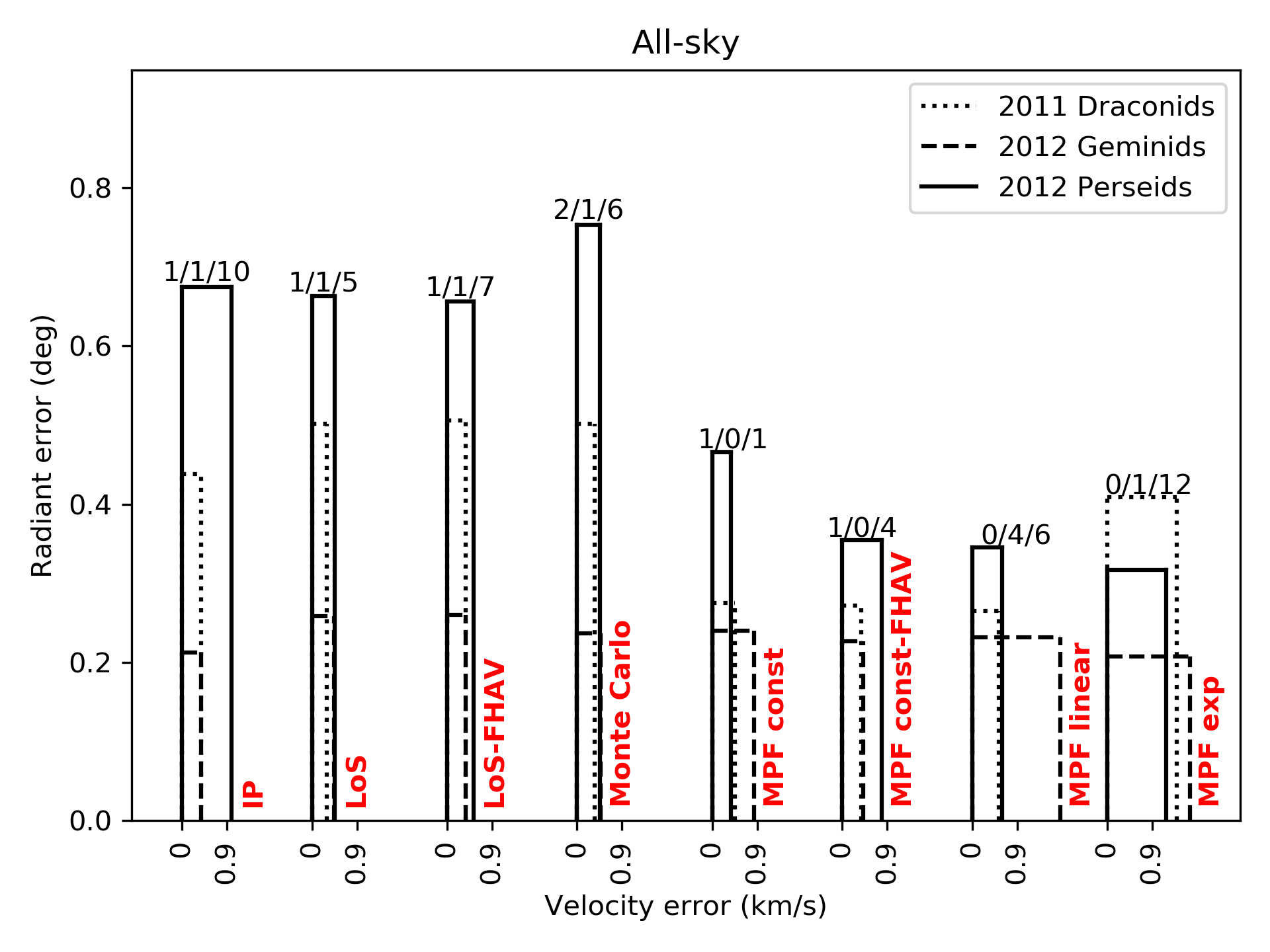}
  \caption{Comparison of geocentric radiant and velocity accuracy for a simulated three station all-sky system for three simulated showers and the various trajectory solvers. The numbers at the top of each vertical bar show the number of failures for a particular method (given in red text) for the Draconids, Geminids and Perseids simulated, respectively.}
  \label{fig:allsky_solver_comparison}
\end{figure}

\begin{figure*}
    \includegraphics[width=.5\linewidth]{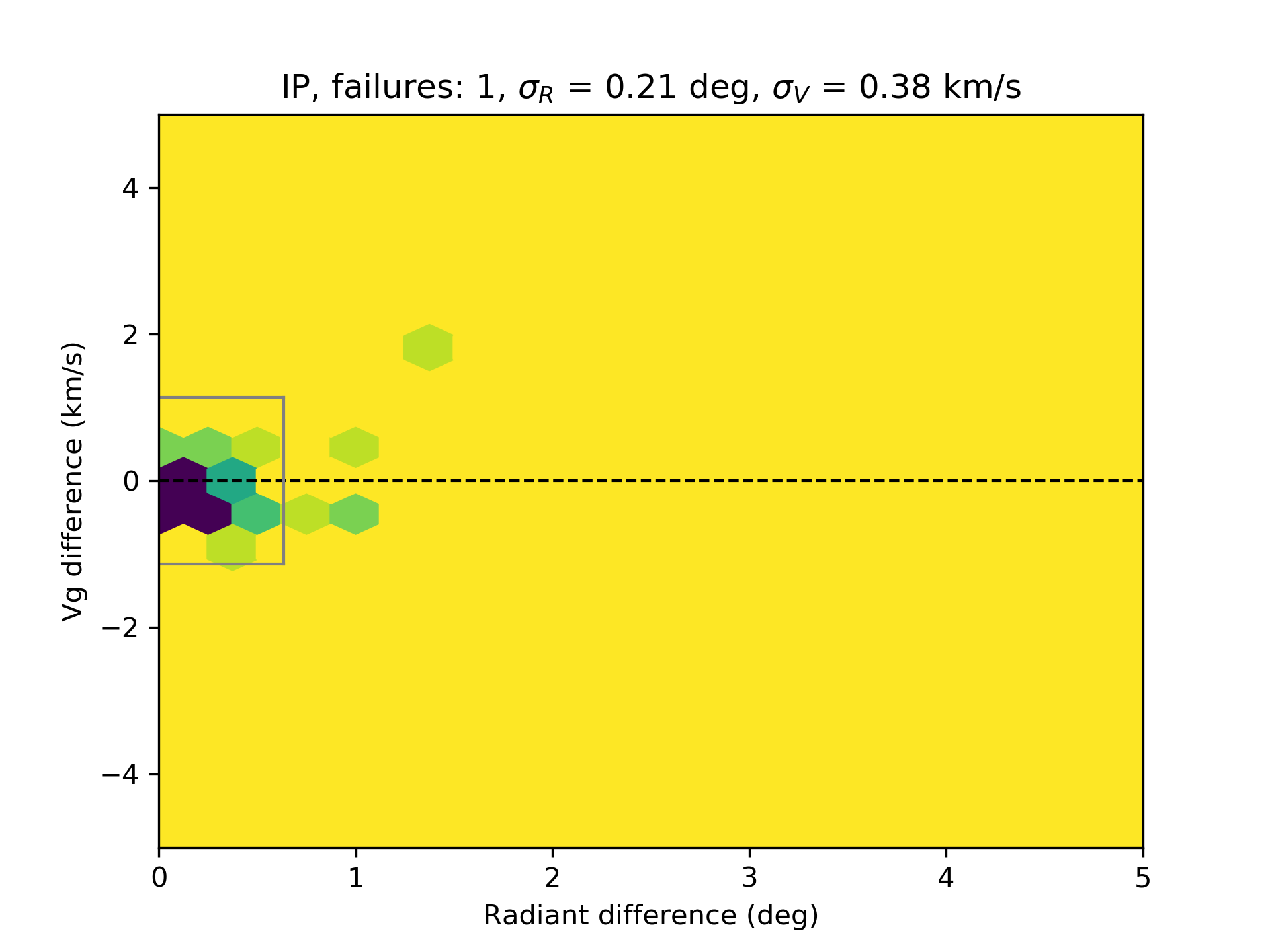}\hfill
    \includegraphics[width=.5\linewidth]{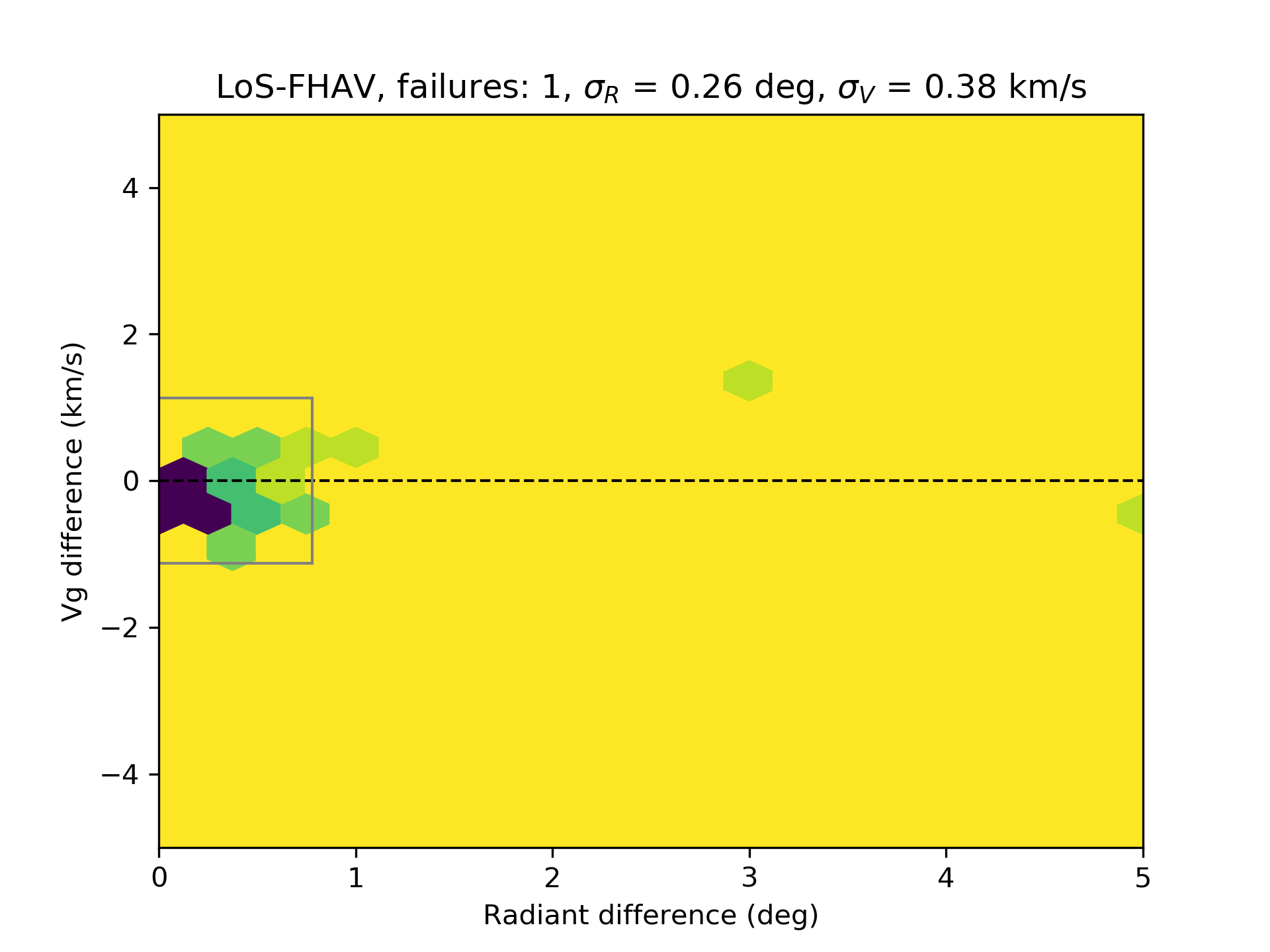}
    \includegraphics[width=.5\linewidth]{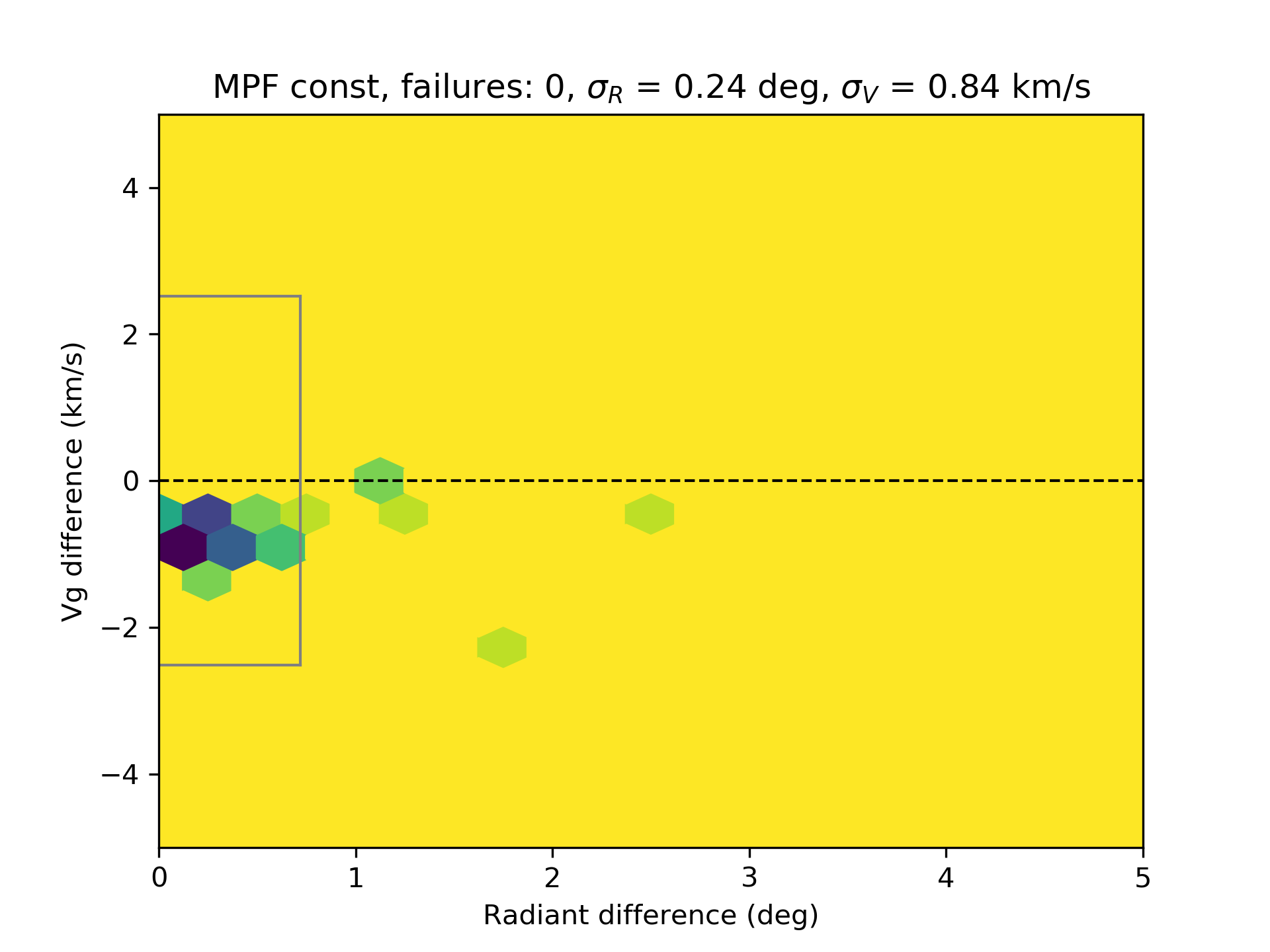}\hfill
    \includegraphics[width=.5\linewidth]{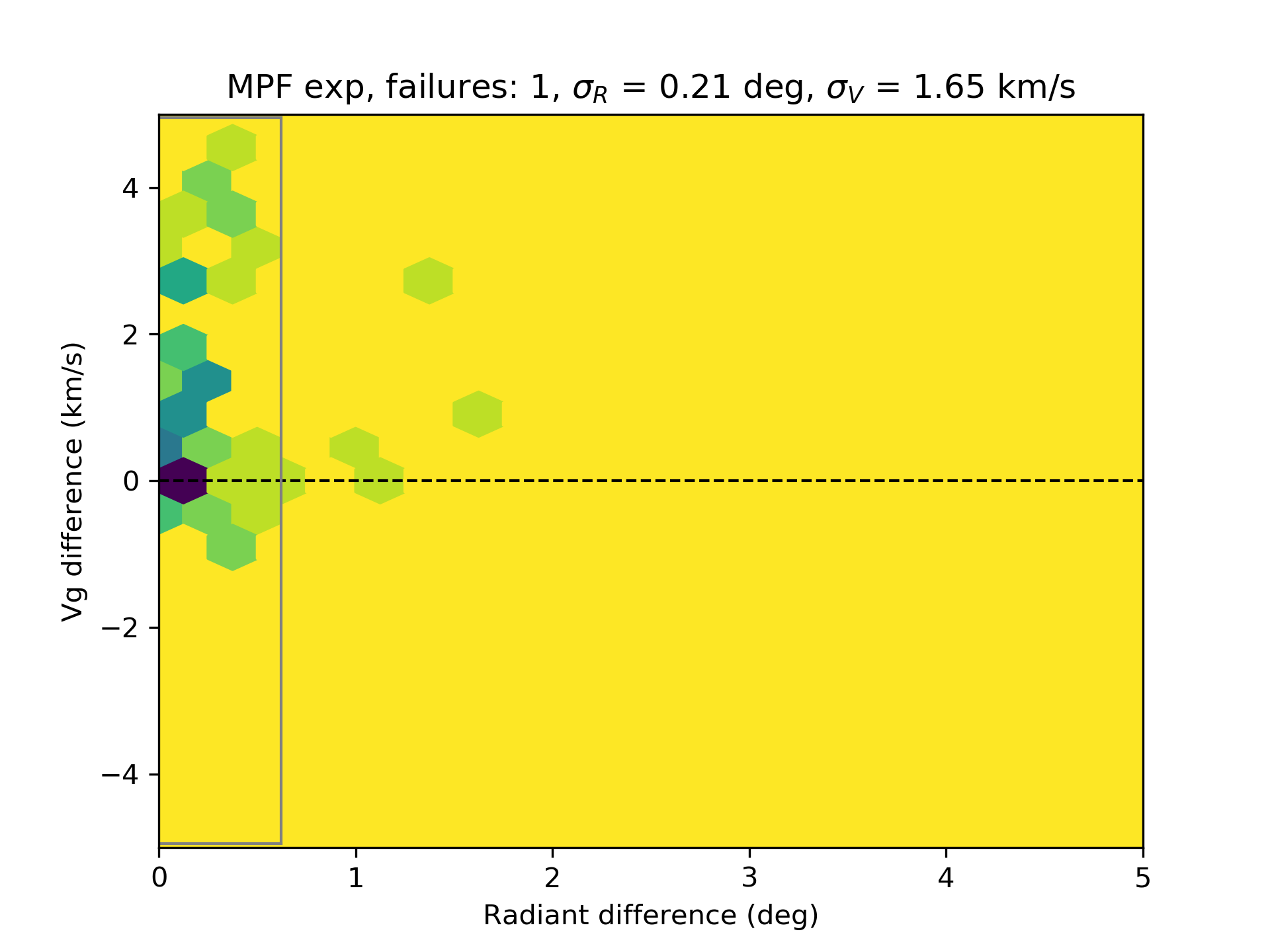}
    \caption{Accuracy of geocentric radiants for the Geminids simulated for all-sky systems shown as a density plot (darker is denser). Upper left: Intersecting planes. Upper right: Lines of sight, initial velocity estimated as the average velocity of the first half of the trajectory. Bottom left: Multi-parameter fit, constant velocity model. Bottom right: Multi-parameter fit, exponential velocity model.}
    \label{fig:allsky_solver_selection}
\end{figure*}

Table \ref{tab:allsky_solvers_performance} lists the accuracy of geocentric radiants computed for the simulated showers using different methods of meteor trajectory estimation for a three station all-sky system. Figure \ref{fig:allsky_solver_comparison} is a visualization of the values in the table. The numbers above the vertical bars for each solver represent the failure rate (out of 100) for the Draconids, the Geminids and the Perseids (in that order). Figure \ref{fig:allsky_solver_selection} shows the distribution of radiant and velocity errors as 2D hexbin histograms for a selection of solvers applied to simulated Geminid data. The increasing bin count is color coded with increasingly darker colors. The gray boxes show the $3\sigma$ standard deviation.

From figure \ref{fig:allsky_solver_comparison} it is apparent that the IP and LoS methods achieve decent radiant accuracy, but tend to have a larger number of failures for faster meteors (Perseids). Moreover, the estimated velocity accuracy decreases with meteor speed, a direct consequence of the smaller number of data points on which the velocity can be estimated for these methods. The Monte Carlo method does not provide any significant increase in accuracy as expected, as the limited precision does not produce useful lag measurements. 

Figure \ref{fig:allsky_solver_selection} shows that the MPF exponential velocity model tends to overestimate velocities at infinity \citep[a behaviour of the exponential deceleration function also noticed by][]{pecina1983new} and that it has a noticeably larger number of failures for faster meteoroids than other methods (except the IP). In contrast, the MPF constant velocity model is the most robust all-around solver for all-sky data, simultaneously achieving good radiant accuracy and a low number of failures, even for faster meteors. Note, however, that because the MPF constant velocity model only computes the average velocity, the velocity estimation is precise but not accurate; it is systematically underestimated. The use of the MPF const-FHAV was an attempt to improve on this deficiency by computing the initial velocity as equal to the average velocity of the first half of the trajectory. This works well for the Draconids and the Geminids, but produces a significantly higher error for the faster Perseids, where the reduction in number of measured points leads to a much larger error.

From our simulations, it appears that the optimal operational approach for low resolution (video) all-sky systems would be to adopt the MPF solver with the constant velocity model, plus a separate (empirical) deceleration correction. The expected geocentric radiant error with this solver is around \ang{0.25} (\ang{0.5} for the Perseids) and around \SI{500}{\metre \per \second} in velocity (or \SI{250}{\metre \per \second} if additional compensation for the early, pre-luminous deceleration is included).

\subsection{Moderate FOV systems}

\begin{figure}
  \includegraphics[width=\linewidth]{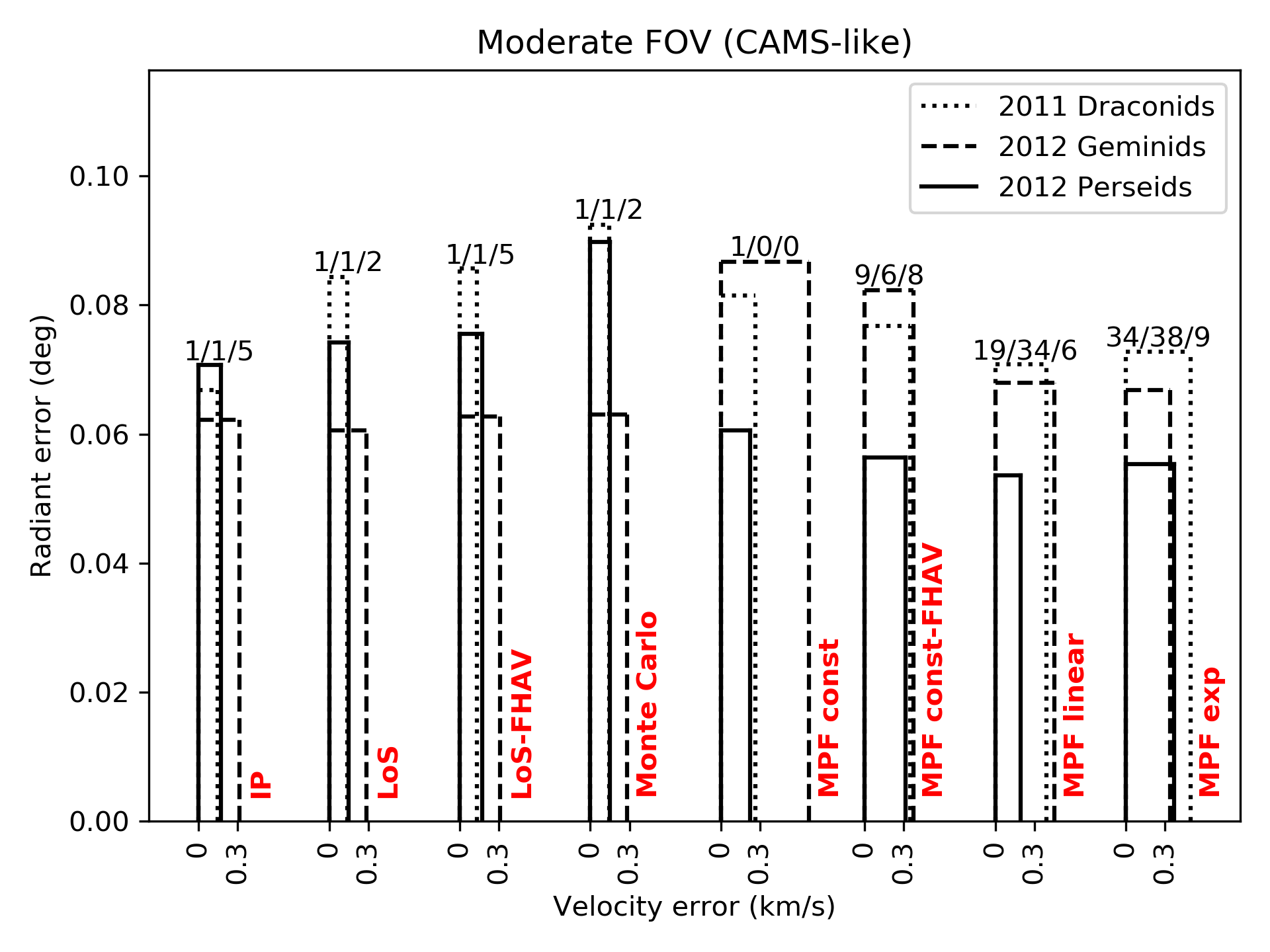}
  \caption{Comparison of the geocentric radiant and velocity accuracy for a simulated CAMS-type system for three showers and various trajectory solvers.}
  \label{fig:cams_solver_comparison}
\end{figure}

Table \ref{tab:cams_solvers_performance} lists the accuracy of geocentric radiants computed for individual showers using different methods of meteor trajectory estimation for simulated meteors detected by a three station CAMS-type optical system. Figure \ref{fig:cams_solver_comparison} shows the visualization of the values in the table.

\begin{table*}
	\caption{Solver performance comparison for the simulated moderate FOV system. The trajectory was included in the statistics if the convergence angle was larger than \ang{10}. F is the number of failures (out of 100), i.e. the number of radiants that were outside the $\Delta_{Rmax} = \ang{1}$, $\Delta_{Vmax} = \SI{1}{\kilo \metre \per \second}$ window.}
    {
	\begin{tabular}{l | c | c | c | c | c | c | c | c | c}
	\hline\hline 
	Solver           & \multicolumn{3}{|c|}{DRA} & \multicolumn{3}{|c|}{GEM} & \multicolumn{3}{|c}{PER} \\
	                 & F & $\sigma_R$ & $\sigma_V$ & F & $\sigma_R$ & $\sigma_V$ & F & $\sigma_R$ & $\sigma_V$ \\
	\hline
IP                &  1 & \ang{0.07} & \SI{0.15}{\kilo \metre \per \second} &  1 & \ang{0.06} & \SI{0.31}{\kilo \metre \per \second} &  5 & \ang{0.07} & \SI{0.17}{\kilo \metre \per \second}\\
LoS               &  1 & \ang{0.09} & \SI{0.17}{\kilo \metre \per \second} &  1 & \ang{0.06} & \SI{0.29}{\kilo \metre \per \second} &  2 & \ang{0.07} & \SI{0.16}{\kilo \metre \per \second}\\
LoS-FHAV          &  1 & \ang{0.09} & \SI{0.13}{\kilo \metre \per \second} &  1 & \ang{0.06} & \SI{0.31}{\kilo \metre \per \second} &  5 & \ang{0.08} & \SI{0.17}{\kilo \metre \per \second}\\
Monte Carlo       &  2 & \ang{0.08} & \SI{0.15}{\kilo \metre \per \second} &  1 & \ang{0.06} & \SI{0.28}{\kilo \metre \per \second} &  2 & \ang{0.09} & \SI{0.15}{\kilo \metre \per \second}\\
MPF const         &  1 & \ang{0.08} & \SI{0.26}{\kilo \metre \per \second} &  0 & \ang{0.09} & \SI{0.67}{\kilo \metre \per \second} &  0 & \ang{0.06} & \SI{0.22}{\kilo \metre \per \second}\\
MPF const-FHAV    &  9 & \ang{0.08} & \SI{0.35}{\kilo \metre \per \second} &  6 & \ang{0.08} & \SI{0.37}{\kilo \metre \per \second} &  8 & \ang{0.06} & \SI{0.32}{\kilo \metre \per \second}\\
MPF linear        & 19 & \ang{0.07} & \SI{0.39}{\kilo \metre \per \second} & 34 & \ang{0.07} & \SI{0.45}{\kilo \metre \per \second} &  6 & \ang{0.05} & \SI{0.19}{\kilo \metre \per \second}\\
MPF exp           & 34 & \ang{0.07} & \SI{0.50}{\kilo \metre \per \second} & 38 & \ang{0.07} & \SI{0.34}{\kilo \metre \per \second} &  9 & \ang{0.06} & \SI{0.37}{\kilo \metre \per \second}\\
	\hline 
	\end{tabular}
	}
	\label{tab:cams_solvers_performance}
\end{table*}

The situation is more complex than was the case for the all-sky system. For convergence angles \ang{>10}, the best results are produced by the classical intersecting planes and the lines of sight solvers, while the solvers which include kinematics perform either marginally worse in the case of the Monte Carlo solver, or significantly worse in the case of multi-parameter fit methods. For all solvers, events which led to a solution within our acceptance window correspond to expected radiant errors around \ang{0.1}. The velocity error is around \SI{200}{\metre \per \second} (around \SI{100}{\metre \per \second} after deceleration correction) for the better solvers. The exception are the Geminids which have a factor of 2 larger velocity uncertainties. They penetrate deeper into the atmosphere due to their asteroidal composition, and decelerate more, which leads to a larger underestimation of the initial velocity.

\begin{figure*}
  \includegraphics[width=.5\linewidth]{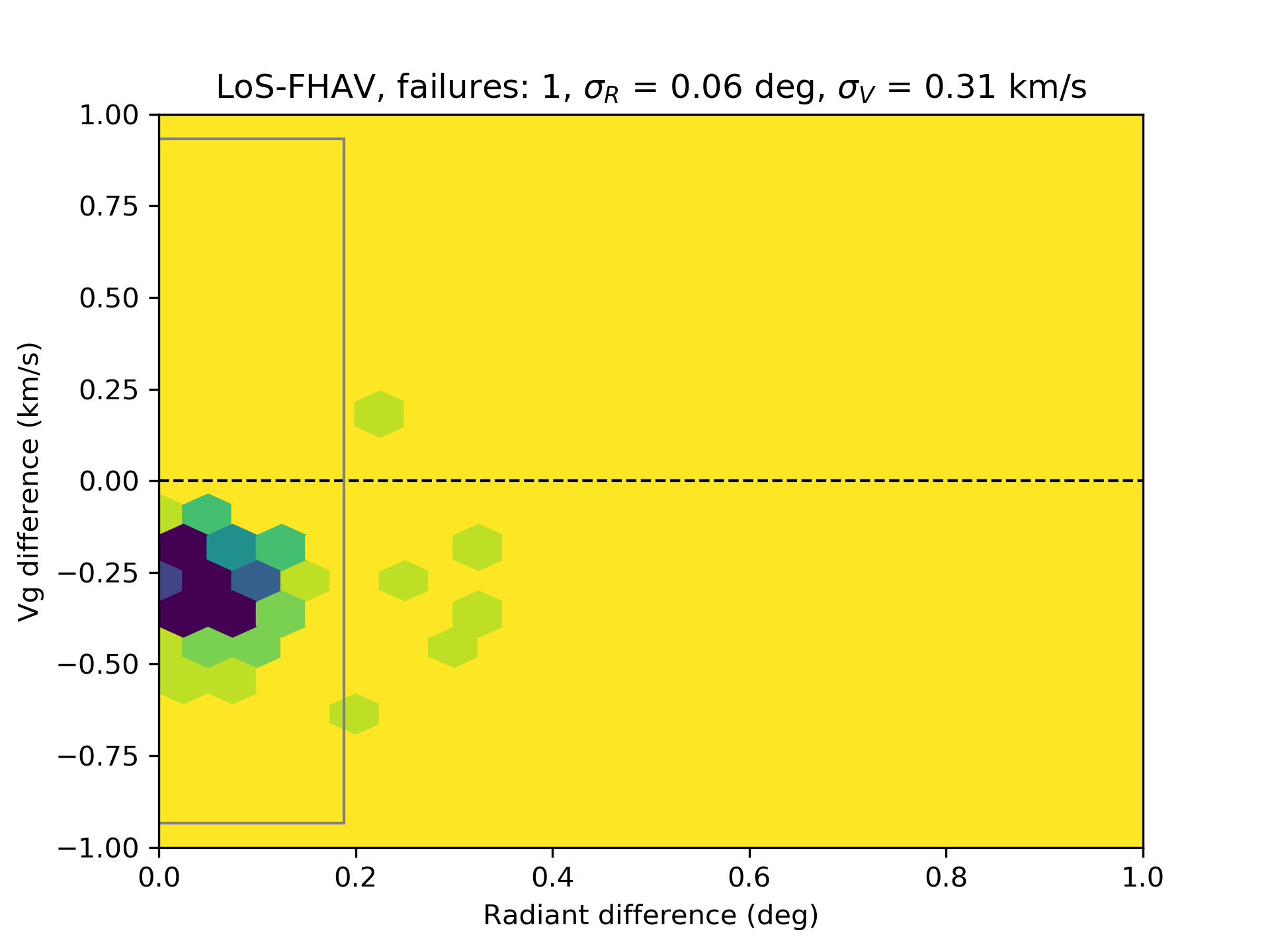}\hfill
  \includegraphics[width=.5\linewidth]{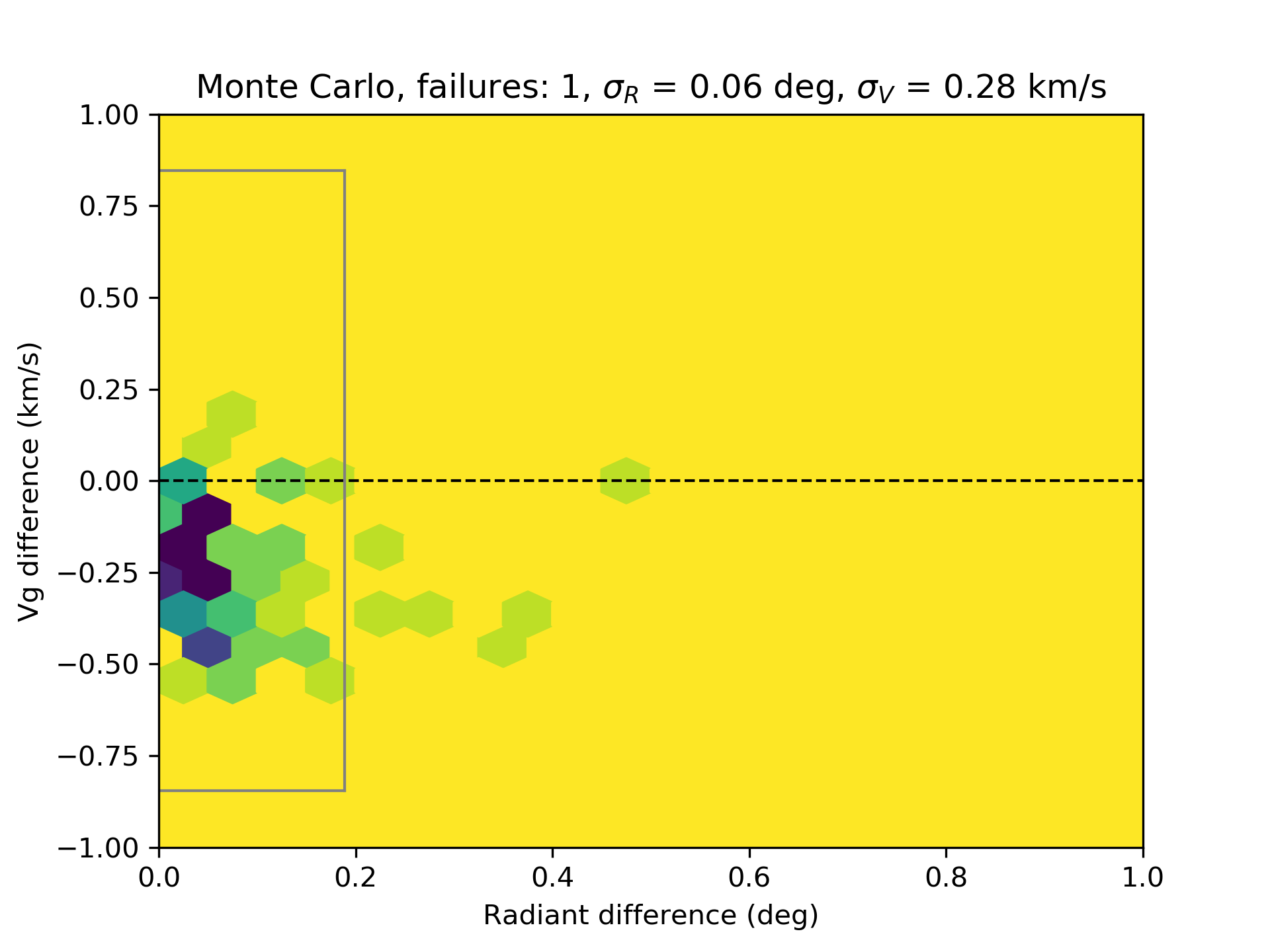}
  \includegraphics[width=.5\linewidth]{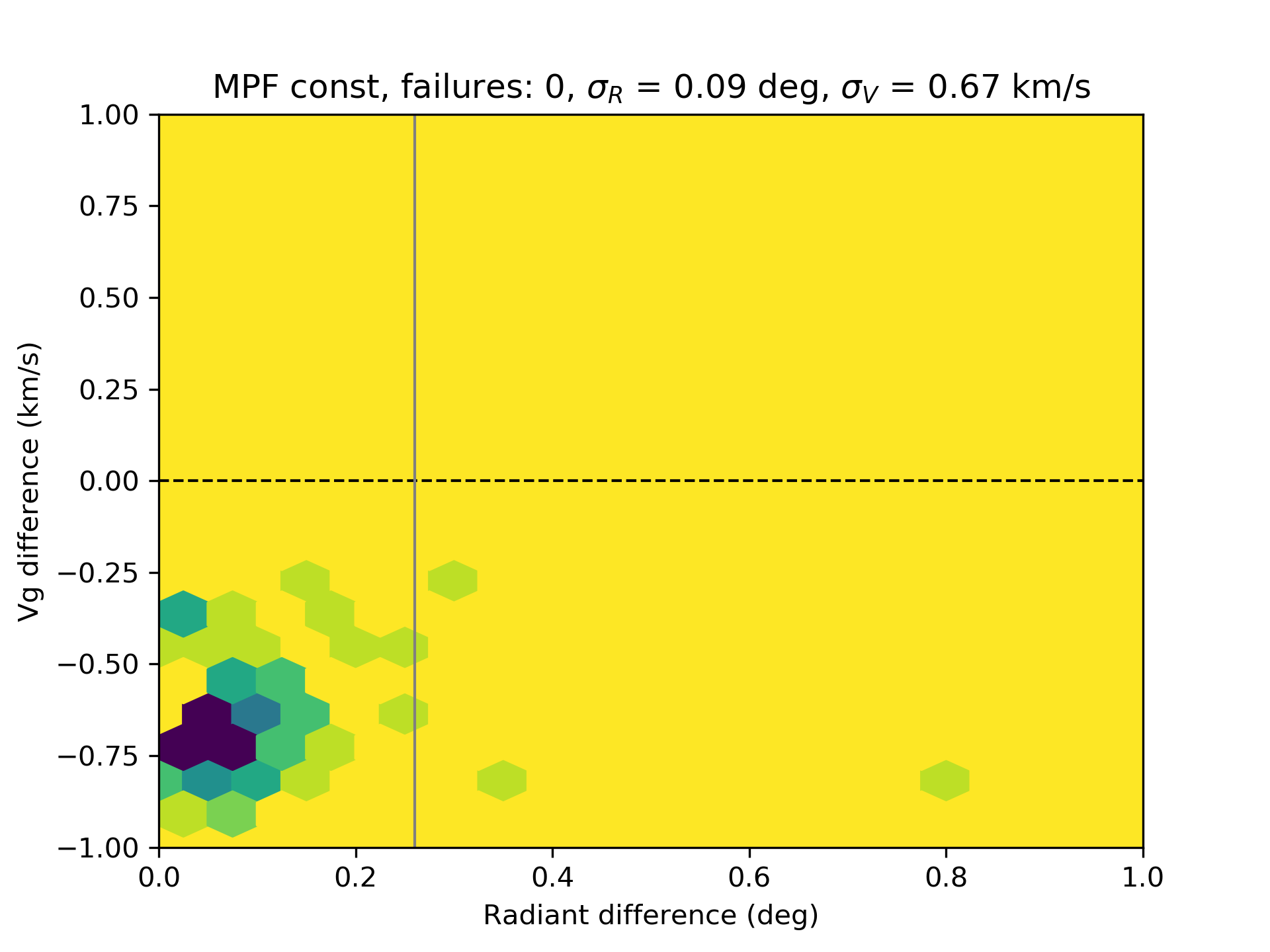}\hfill
  \includegraphics[width=.5\linewidth]{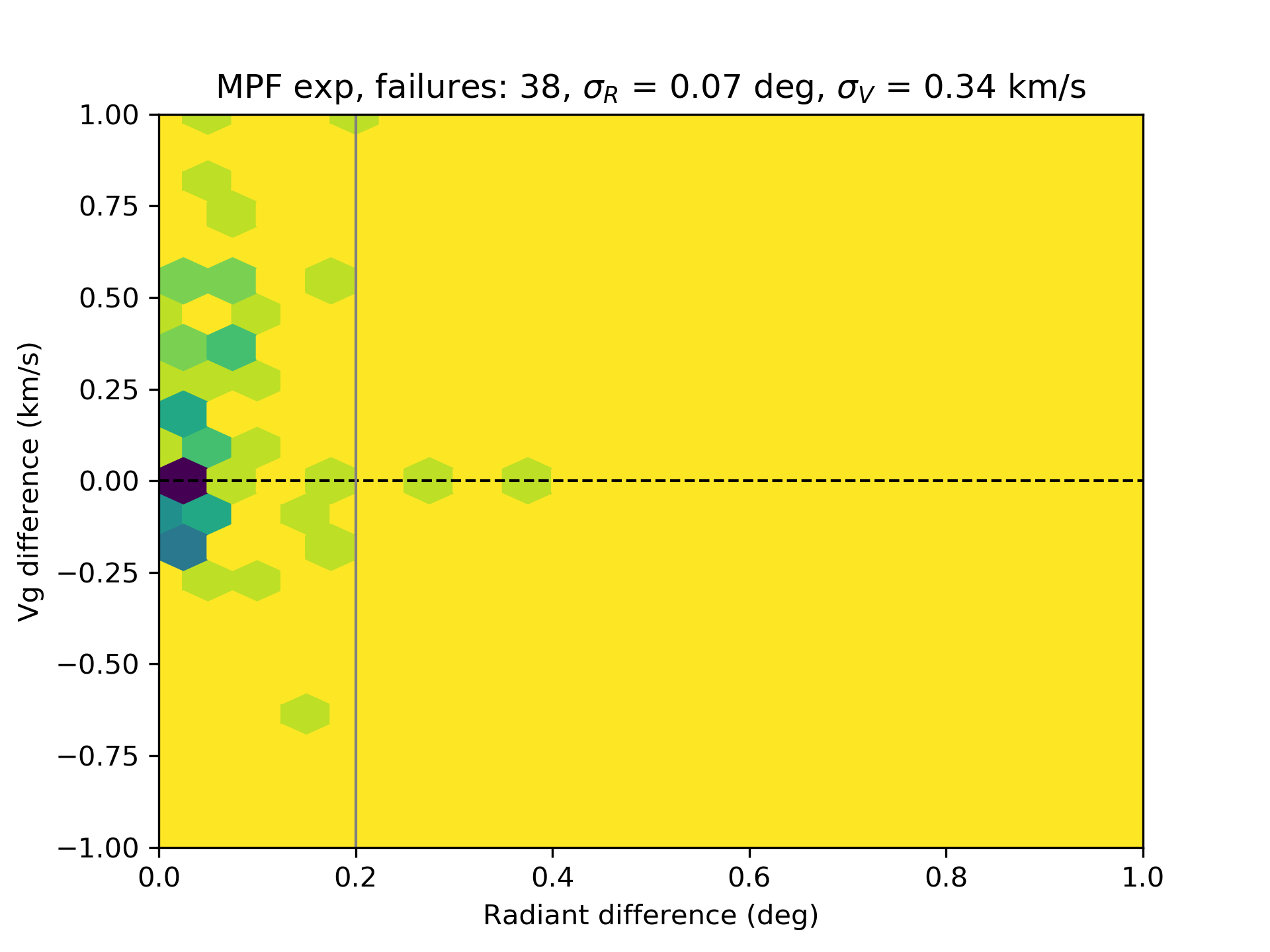}
  \caption{Accuracy of geocentric radiants for the Geminids simulated for CAMS-like systems. Upper left: Lines of sight, initial velocity estimated as the average velocity of the first half of the trajectory. Upper right: Monte Carlo. Bottom left: Multi-parameter fit, constant velocity model. Bottom right: Multi-parameter fit, exponential velocity model.}
  \label{fig:cams_solver_selection}
\end{figure*}

We emphasize that the MPF methods sometimes do produce better estimates of the radiant, which is consistent with \cite{gural2012solver} who only investigated the precision of the radiant position for various solvers. On the other hand, MPF-based velocity estimates are consistently worse by a factor of 2 or more when compared to other methods, as shown in figure \ref{fig:cams_solver_selection}. The MPF method with the constant velocity model does produce robust (precise) solutions, but still requires either correcting for the deceleration or an alternate way of computing the initial velocity (and improving accuracy).

Computing the initial velocity as the average of the first half (MPF const-FHAV) does not result in an improvement but causes an even larger spread in the estimated velocities. Furthermore, the MPF method with the exponential deceleration model produces a high failure rate for this type of data as well, predominantly due to the overestimation of the initial velocity. On the other hand, the radiant estimation was as robust as the other solvers, for events which met our acceptance criteria. This finding is worrisome as this velocity model was used by the CAMS network \citep{jenniskens2016established}. Our simulations suggest that initial velocities obtained with MPF-exp for moderate field of view systems should ideally be compared to other solvers before acceptance. 

\subsection{CAMO system}

\begin{figure}
  \includegraphics[width=\linewidth]{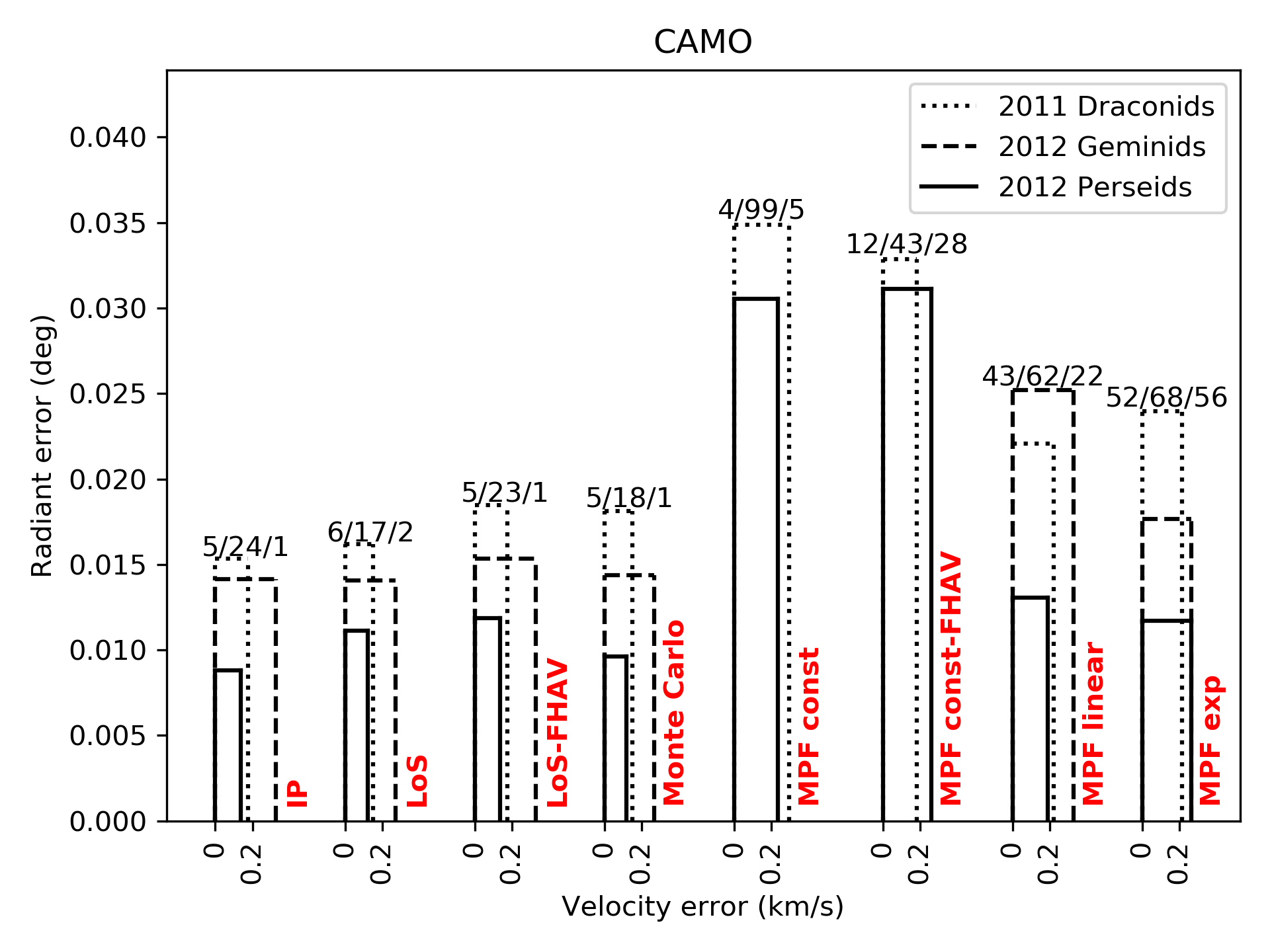}
  \caption{Comparison of geocentric radiant and velocity accuracy for the simulated CAMO system for three simulated showers and various trajectory solvers.}
  \label{fig:camo_solver_comparison}
\end{figure}

\begin{table*}
	\caption{Solver performance comparison for the simulated CAMO-like optical system. A simulated trajectory was included in the final statistics if the convergence angle was larger than \ang{1}. F is the number of failures (out of 100), i.e. the number of radiants that were outside the $\Delta_{Rmax} = \ang{0.5}$, $\Delta_{Vmax} = \SI{0.5}{\kilo \metre \per \second}$ window.}
    {
	\begin{tabular}{l | c | c | c | c | c | c | c | c | c}
	\hline\hline 
	Solver           & \multicolumn{3}{|c|}{DRA} & \multicolumn{3}{|c|}{GEM} & \multicolumn{3}{|c}{PER} \\
	                 & F & $\sigma_R$ & $\sigma_V$ & F & $\sigma_R$ & $\sigma_V$ & F & $\sigma_R$ & $\sigma_V$ \\
	\hline
IP                &  5 & \ang{0.02} & \SI{0.18}{\kilo \metre \per \second} & 24 & \ang{0.01} & \SI{0.33}{\kilo \metre \per \second} &  1 & \ang{0.01} & \SI{0.14}{\kilo \metre \per \second}\\
LoS               &  8 & \ang{0.02} & \SI{0.15}{\kilo \metre \per \second} & 17 & \ang{0.01} & \SI{0.27}{\kilo \metre \per \second} &  1 & \ang{0.01} & \SI{0.12}{\kilo \metre \per \second}\\
LoS-FHAV          &  5 & \ang{0.02} & \SI{0.17}{\kilo \metre \per \second} & 23 & \ang{0.02} & \SI{0.33}{\kilo \metre \per \second} &  1 & \ang{0.01} & \SI{0.14}{\kilo \metre \per \second}\\
Monte Carlo       &  6 & \ang{0.02} & \SI{0.15}{\kilo \metre \per \second} & 18 & \ang{0.01} & \SI{0.27}{\kilo \metre \per \second} &  1 & \ang{0.01} & \SI{0.11}{\kilo \metre \per \second}\\
MPF const         &  4 & \ang{0.03} & \SI{0.29}{\kilo \metre \per \second} & 99 & \ang{0.31} & \SI{0.47}{\kilo \metre \per \second} &  5 & \ang{0.03} & \SI{0.23}{\kilo \metre \per \second}\\
MPF const-FHAV    & 12 & \ang{0.03} & \SI{0.18}{\kilo \metre \per \second} & 43 & \ang{0.17} & \SI{0.35}{\kilo \metre \per \second} & 28 & \ang{0.03} & \SI{0.26}{\kilo \metre \per \second}\\
MPF linear        & 43 & \ang{0.02} & \SI{0.22}{\kilo \metre \per \second} & 62 & \ang{0.03} & \SI{0.33}{\kilo \metre \per \second} & 22 & \ang{0.01} & \SI{0.19}{\kilo \metre \per \second}\\
MPF exp           & 52 & \ang{0.02} & \SI{0.21}{\kilo \metre \per \second} & 68 & \ang{0.02} & \SI{0.27}{\kilo \metre \per \second} & 56 & \ang{0.01} & \SI{0.26}{\kilo \metre \per \second}\\

	\hline 
	\end{tabular}
	}
	\label{tab:camo_solvers_performance}
\end{table*}

Table \ref{tab:camo_solvers_performance} lists the accuracy of geocentric radiants for our three modelled showers using different methods of meteor trajectory estimation applied to simulated CAMO data. Figure \ref{fig:camo_solver_comparison} shows a visualization of the values in the table. Note that the Geminids are missing from the graph for the MPF const method, as most Geminid velocities estimated with that method were outside the \SI{0.5}{\kilo \metre \per \second} threshold due to the larger deceleration of these asteroidal meteoroids. Thus, in this section we use the Draconids for the comparison between solvers.

\begin{figure*}
  \includegraphics[width=.5\linewidth]{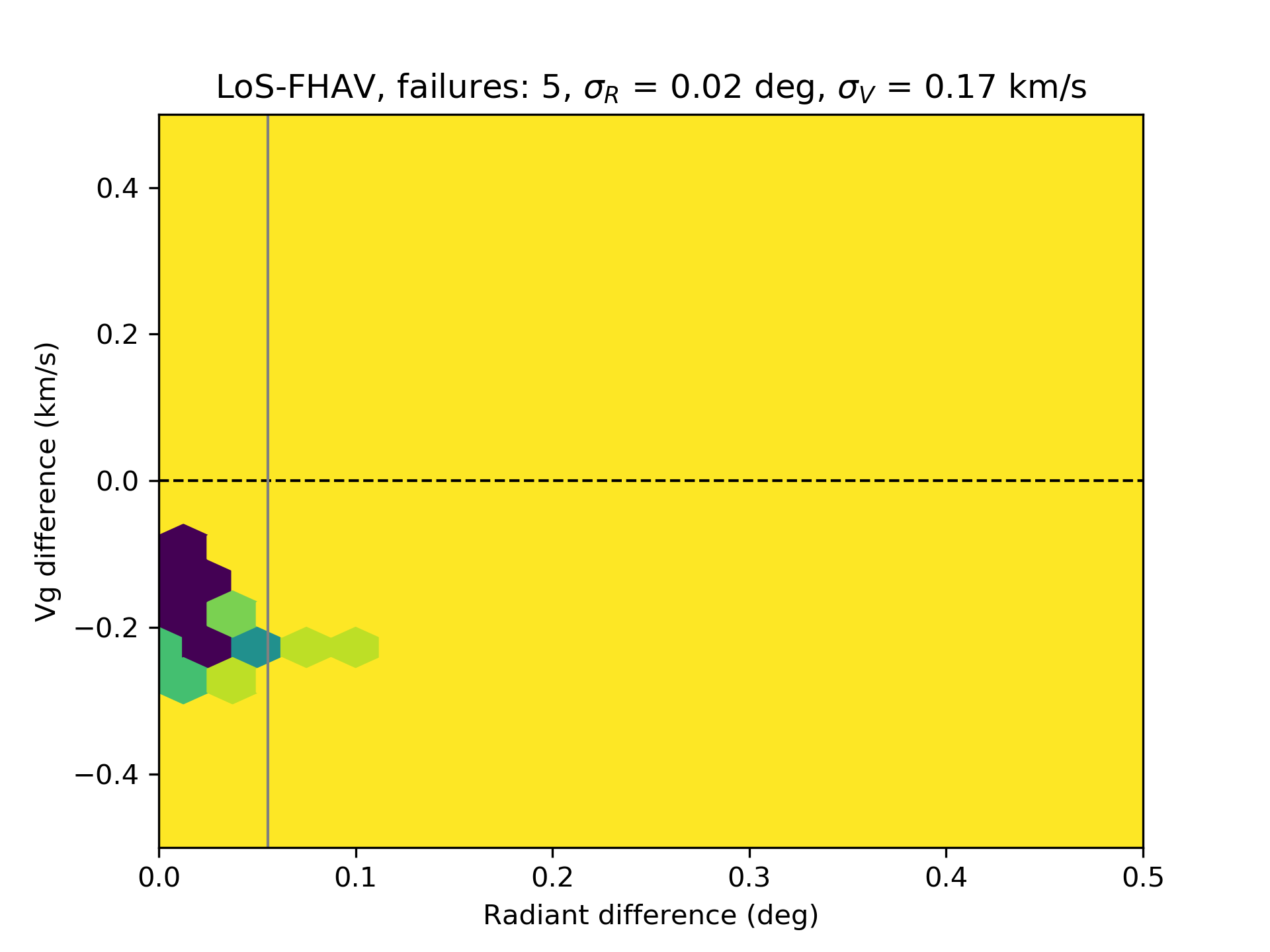}\hfill
  \includegraphics[width=.5\linewidth]{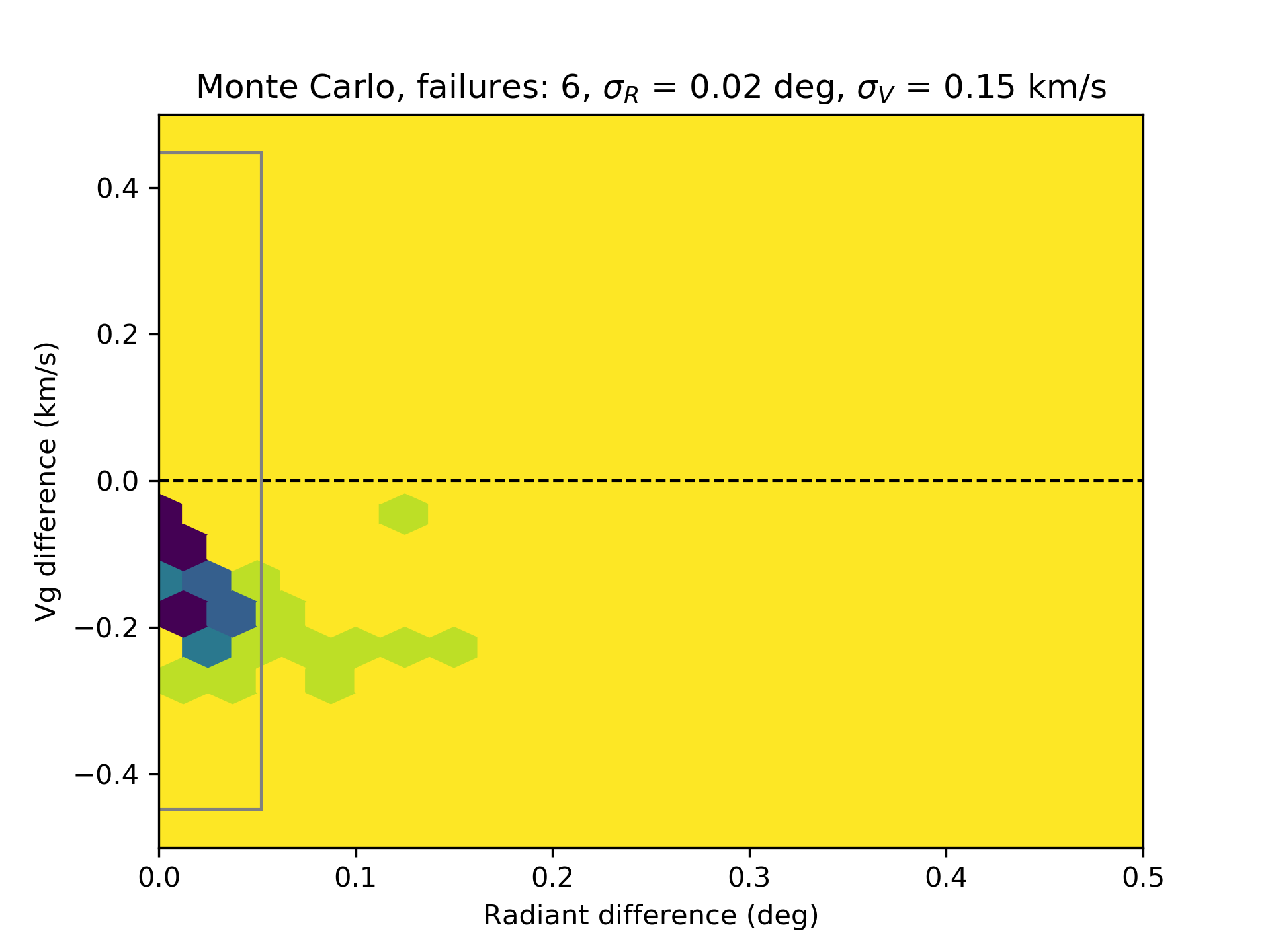}
  \includegraphics[width=.5\linewidth]{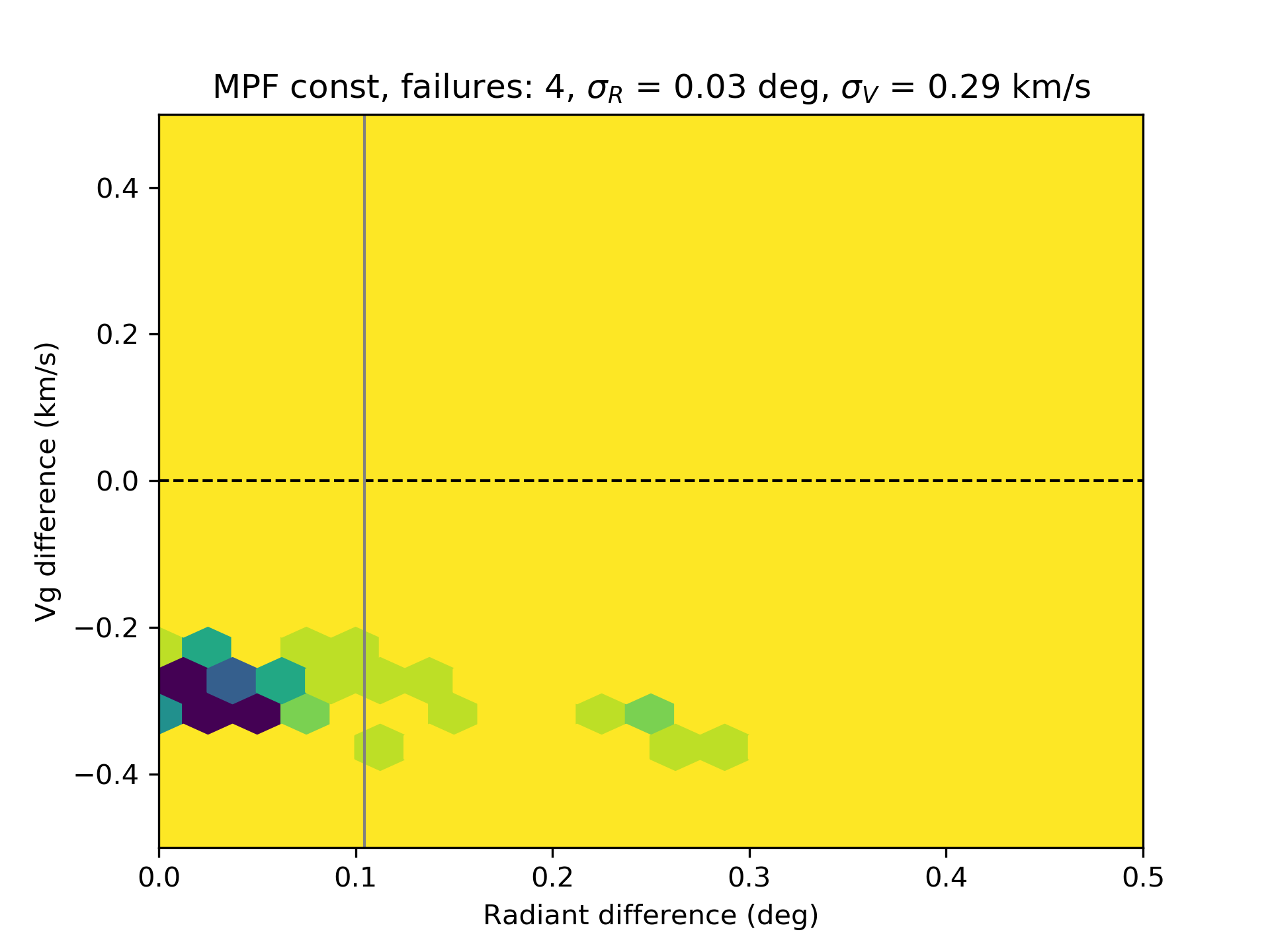}\hfill
  \includegraphics[width=.5\linewidth]{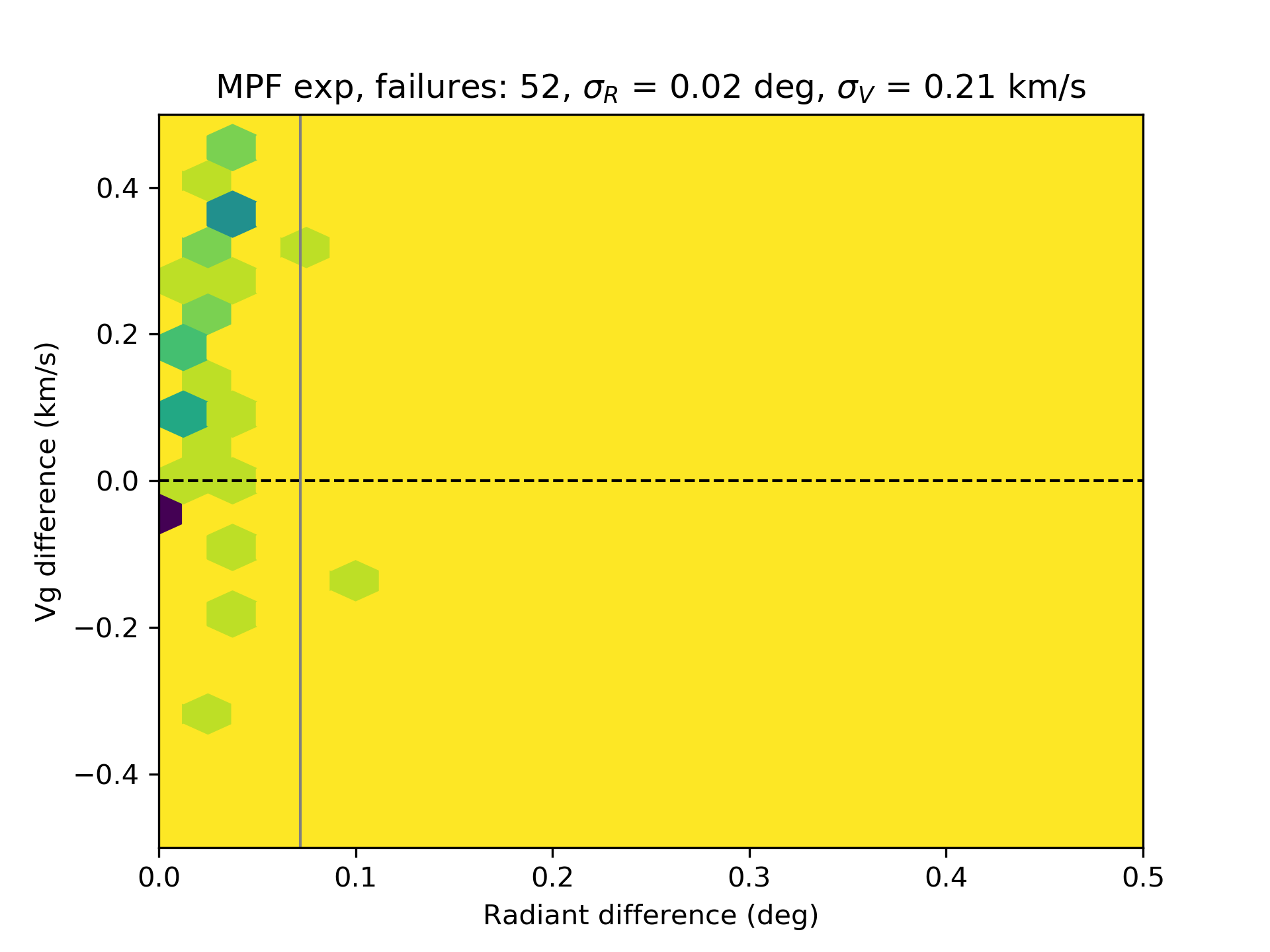}
  \caption{Accuracy of simulated CAMO geocentric radiants for the Draconids. Upper left: Lines of sight, initial velocity estimated as the average velocity of the first half of the trajectory. Upper right: Monte Carlo. Bottom left: Multi-parameter fit, constant velocity model. Bottom right: Multi-parameter fit, exponential velocity model.}
  \label{fig:camo_solver_selection}
\end{figure*}

The upper left inset of figure \ref{fig:camo_solver_selection} shows the results obtained using the LoS-FHAV method for the Draconids. Only 5 out of 100 solutions failed. The geocentric velocities are systematically underestimated due to deceleration occurring prior to detection. For this system, the average underestimation is around \SI{200}{\metre \per \second} for cometary, and \SI{300}{\metre \per \second} for asteroidal meteoroids \citep[see][for a complete analysis]{vida2018modeling}. The measurement precision of the initial velocity is much better, around \SI{50}{\metre \per \second}. The accuracy of the radiant estimation is approximately \ang{0.02}.

The upper right inset of figure \ref{fig:camo_solver_selection} shows the results obtained using the Monte Carlo solver. Overall this solver and the LoS solver provided the best precision for CAMO data. Both have very low failure rates; the radiant accuracy was around \ang{0.01} and the geocentric velocity accuracy around \SI{150}{\metre \per \second}. The geocentric velocity was systematically underestimated due to deceleration prior to detection; the accuracy could be improved by applying the correction given in \cite{vida2018modeling}.

The lower left inset of figure \ref{fig:camo_solver_selection} shows simulation results using the MPF method with the constant velocity model for the Draconids. The geocentric velocity was underestimated more than with the LoS-FHAV method, as the initial velocity estimate is very heavily influenced by deceleration. The average difference between the initial velocity and true  velocity for our 100 simulated Draconids was around \SI{300}{\metre \per \second}. This difference drives the error in the radiant, which had a standard deviation among our simulations of \ang{0.04}. For the Geminds, the velocity difference with this method was \SI{> 1}{\kilo \metre \per \second}, which strongly indicates that using average meteor velocities is not suitable for computing orbits of asteroidal meteors.

Finally, the lower right inset of figure \ref{fig:camo_solver_selection} shows results obtained using the MPF solver with the exponential deceleration model. This solver had a very large failure rate, above 50\%. The failure was mostly driven by the overestimation of the initial velocity; in contrast the estimation of the radiant position remained fairly robust.

\subsection{Trajectory solution accuracy as a function of convergence angle} \label{subsec:convergence_angle_study}

The maximum convergence angle between a meteor trajectory and stations is usually used as an indicator for the trajectory quality. \cite{gural2012solver} has shown that the radiant error is dependent on the convergence angle (among other factors). He found that the IP and LoS methods produced on average a factor of 10 increase in radiant error at low ($< \ang{10}$) convergence angles. In that work, the constant velocity MPF method significantly improved the radiant accuracy, and the error for low convergence angles was only a factor of 2 higher as compared to larger angles.  To test for convergence angle sensitivity amongst solvers, we generated 1000 synthetic Geminids with our ablation model for all three systems. We divided the range of convergence angles into 30 bins of equal numbers of data points; thus 1000 simulations were needed for better statistics. 

\begin{figure}
  \includegraphics[width=\linewidth]{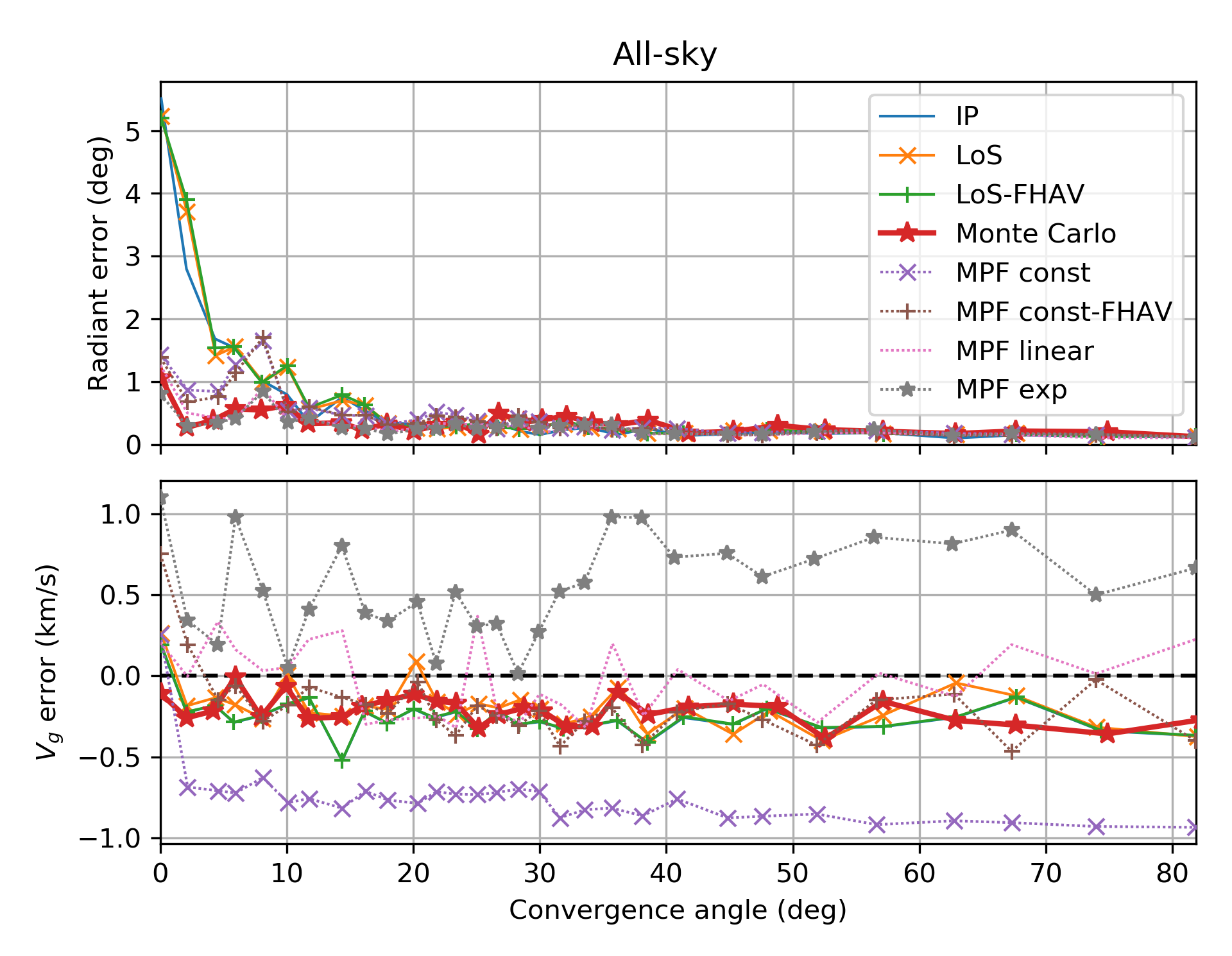}
  \caption{Radiant and velocity error as a function of convergence angle for 1000 Geminids simulated for an all-sky (SOMN-like) system.}
  \label{fig:somn_convergence_angle}
\end{figure}

\begin{figure*}
  \includegraphics[width=\linewidth]{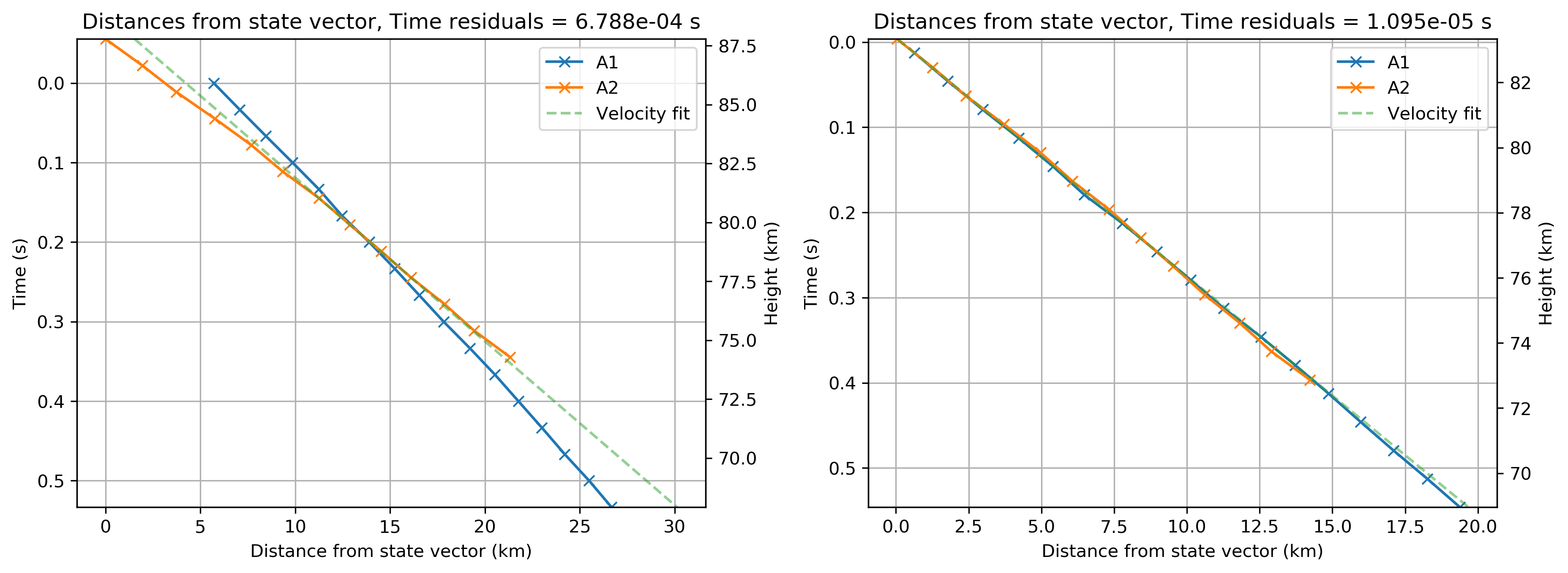}
  \caption{Length as a function of time of a low $Q_c$ Geminid estimated with the LoS method (left) and the Monte Carlo method (right).}
  \label{fig:low_qc_length_comparison}
\end{figure*}

Figure \ref{fig:somn_convergence_angle} shows radiant and velocity errors versus the convergence angle, $Q_c$, for a simulated all-sky (SOMN-like) system. Only two stations (A1 and A2) were used for the convergence angle analysis - when a third station is included, all maximum convergence angles are usually $> \ang{30}$. The plotted data shows the median error value in every bin which contains $\sim 33$ meteors. The geometrical IP and LoS methods produce errors on the order of degrees for $Q_c < \ang{15}$, while the Monte Carlo and MPF methods restrict the radiant error below \ang{1} even for very low values of $Q_c$. Although the deceleration is not directly observable for such all-sky systems, the estimated speeds at different stations at low convergence angles do not match when geometrical methods are used. The Monte Carlo and MPF methods, in contrast, are designed to ``find'' the solution which satisfies both the spatial and dynamical constraints. As an example, figure \ref{fig:low_qc_length_comparison} shows length vs. time of a low $Q_c$ synthetic Geminid estimated by the LoS method (left inset) and the Monte Carlo method (right inset). Note that in all the convergence angle plots, the IP and LOS-FHAV use the same approach to compute speed and hence overlap exactly in the speed error (bottom) plots and are not separately observable in these figures.

\begin{figure}
  \includegraphics[width=\linewidth]{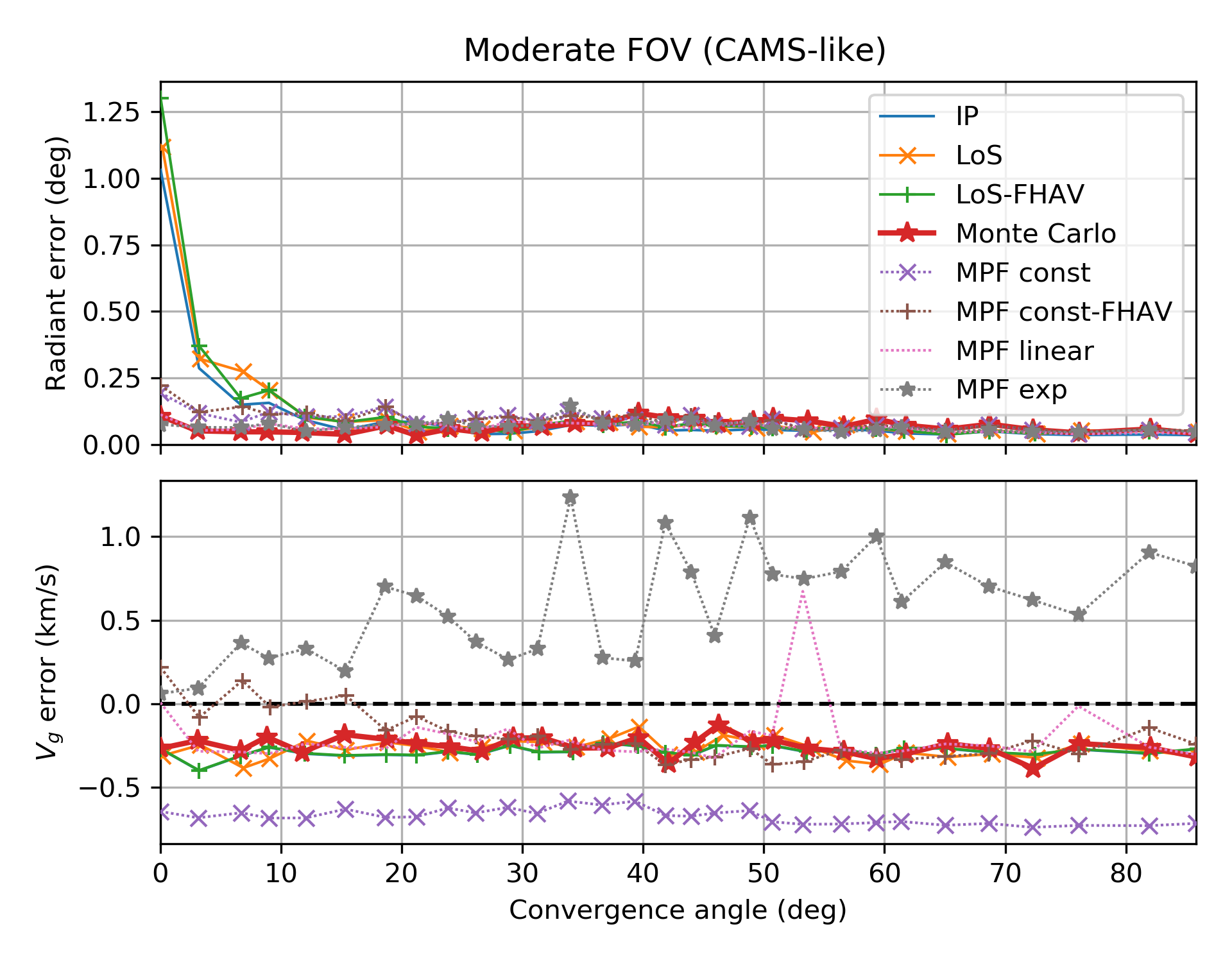}
  \caption{Radiant and velocity error vs convergence angle for 1000 Geminids simulated for a moderate field of view (CAMS-like) system.}
  \label{fig:cams_convergence_angle}
\end{figure}

Next, we investigated a moderate field of view (CAMS-like) system. As with the all-sky system, only two stations were used in this analysis, as three-station solutions always have large convergence angles. As seen in figure \ref{fig:cams_convergence_angle}, all trajectory solvers have a similar radiant error of $\ang{\sim 0.1}$ for $Q_c > \ang{10}$. For smaller convergence angles, the IP and LoS solvers produce errors on the order of $\ang{1}$. On the other hand, the MPF and Monte Carlo solvers produce robust radiant solutions throughout. The velocity error is less strongly correlated with the convergence angle, but is completely dominated by solver-specific biases. In particular, the MPF-exp overestimates the initial velocity across all convergence angles, a product of the poorly conditioned convergence of this kinematic model. We are roughly able to reproduce the results of \cite{jenniskens2011cams}, where they found that geometrical methods work well for $Q_c \ang{> 25}$, and that MPF methods produce good radiant convergences even down to $Q_c  \ang{\sim 2}$.

\begin{figure}
  \includegraphics[width=\linewidth]{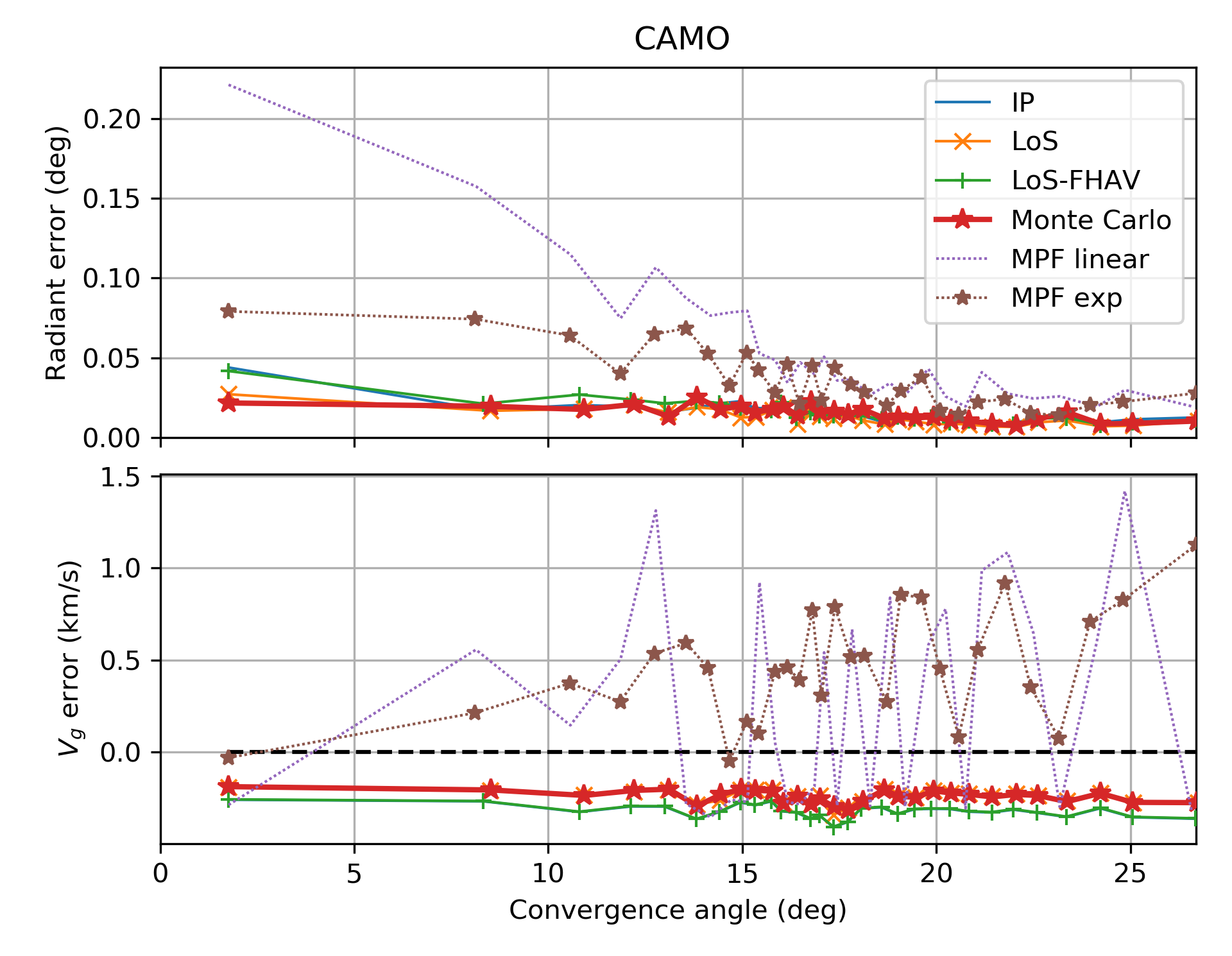}
  \caption{Radiant and velocity error as a function of convergence angle for 1000 Geminids simulated for the CAMO system.}
  \label{fig:camo_convergence_angle}
\end{figure}

Finally, figure \ref{fig:camo_convergence_angle} shows the dependence of radiant and velocity errors on the convergence angle for the CAMO system. The results of the MPF method with the constant velocity model are excluded because it produces very large errors since it fails to model the deceleration visible at the fine angular resolution of a CAMO-like system. Due to the limited geometry of the CAMO system, the maximum expected convergence angle is less than $\ang{30}$, but the solutions using the IP, LoS and Monte Carlo solvers are very stable even down to $Q_c \ang{\sim 1}$. The poorer performance of the MPF-methods for this system, particularly in reconstructing initial speed, likely reflect the high precision of CAMO which requires a more physical kinematic model than the empirical models used in the MPF approach.

\subsection{The 2015 Taurid outburst - high-precision all-sky observations} \label{subsec:2015taurids}

In 2015 the Taurids displayed an activity outburst due to Earth encountering a resonant meteoroid swarm locked in a 7:2 resonance with Jupiter \citep{asher1993extraterrestrial, spurny2017discovery}. Securing accurate observations of fireballs for this resonant branch for the first time is significant for planetary defence as the branch was also shown to contain asteroids on the order of several 100s of meters in size. As the Earth encounters this swarm at regular intervals it may be a major contributor to the overall  impact hazard and supports some elements of the proposed coherent catastrophism theory \citep{asher1994coherent}.

\cite{spurny2017discovery} attribute the discovery of the Taurid branch, linked to the swarm return in 2015, to the precision measurements made possible by their high-resolution all-sky digital cameras and careful manual reduction of the data. Figure 14 in \cite{spurny2017discovery} shows that all meteoroids in the resonant swarm reside in a very narrow range of semi-major axes \cite[the extent of the 7:2 resonance given by][]{asher1993extraterrestrial}. This is arguably  the strongest evidence yet published for the real existence of the branch and swarm as the semi-major axis is very sensitive to measurement errors in meteor velocity. \cite{olech2017enhanced} also noticed a possible connection to the 7:2 resonance, but their results were not as conclusive due to the lower measurement accuracy. 

Data for the outburst have only been published for fireball-sized meteoroids and we focus on simulating these for comparison to measurements published in \cite{spurny2017discovery}. In this section we investigate the precision needed to make a discovery of this nature and discuss the limits of low-resolution all-sky systems.

We used the parameters of the Taurid meteoroid stream resonant branch as given in \cite{spurny2017discovery} and simulated 100 meteors from the branch as they would be detected by an SOMN-like all-sky system. We only simulated meteors with the semi-major axis inside the narrow region of the 7:2 resonance which spans 0.05 AU.

\begin{figure*}
  \includegraphics[width=\linewidth]{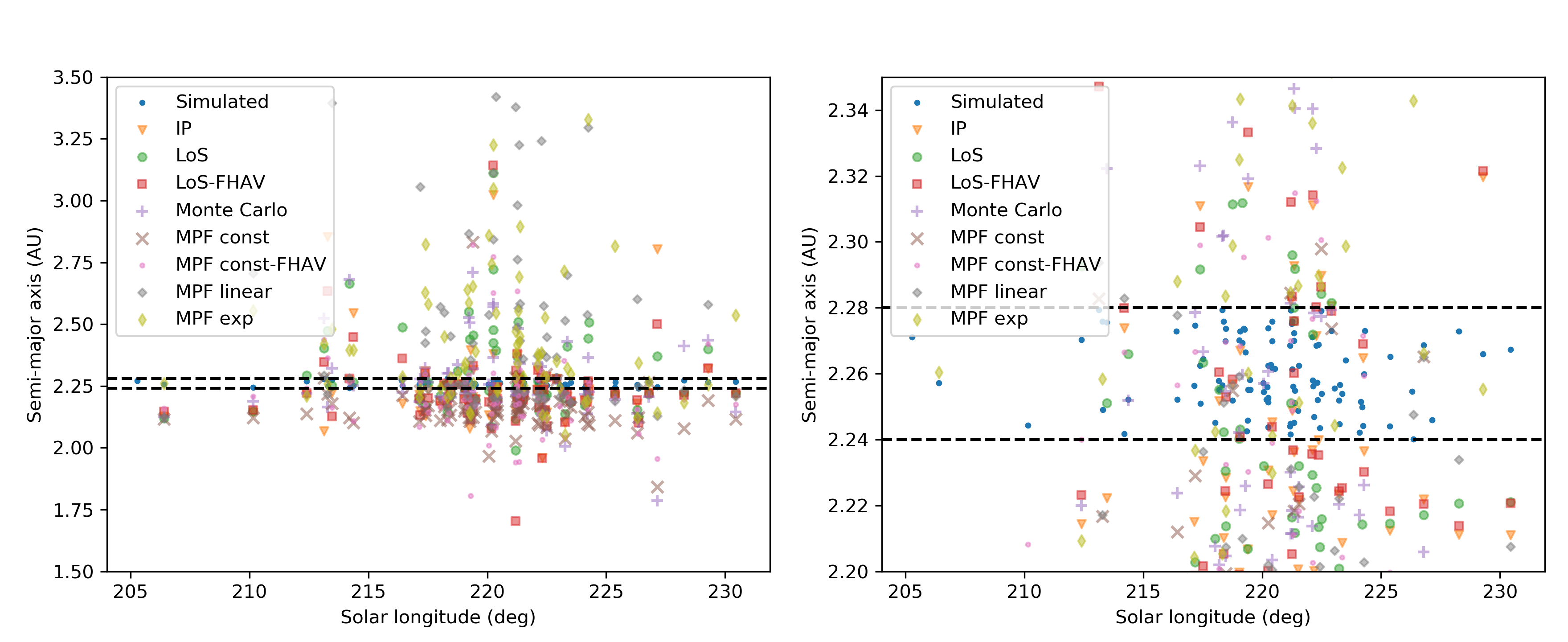}
  \caption{The semi-major axes of simulated 2015 Taurids from the resonant branch (blue dots bound within black dashed lines) and comparison with values estimated using different trajectory solvers. The right plot shows a narrower range of semi-major axes.}
  \label{fig:2015taurids_lasun_a_comparison}
\end{figure*}

Figure \ref{fig:2015taurids_lasun_a_comparison} shows the comparison between the simulated semi-major axes and computed values using various trajectory solvers for a three station SOMN-like system. As can be seen, the observed scatter in $a$ is too large to detect the branch with any solver using such low-quality data. This shows that existing low-resolution all-sky systems (circa 2019) have limited utility for any orbital determination (on a per camera basis) requiring high precision but are better suited to shower flux estimation or meteorite recovery. Deployment of higher resolution all-sky systems \citep{spurny2006automation, devillepoix2018observation} which can achieve a radiant precision on the order of 1 arc minute are clearly preferable for any orbit measurements requiring high accuracy.

\subsection{Influence of gravity on trajectories of long-duration fireballs} \label{subsec:long_fireballs}

The orbits of meteorite-dropping fireballs are of special interest as knowing their velocities allows a statistical estimate of the source region in the main asteroid belt and linkage to possible parent bodies \citep{granvik2018identification}. Because these fireballs may remain luminous for long durations, often 5 or more seconds, their trajectories can be precisely estimated using data even from low-precision all-sky systems. This is due to the large number of observed points available in the data reduction. On the other hand, these events experience the largest gravitational bending of the trajectory, e.g. after \SI{5}{\second} the drop due to gravity is on the order of \SI{100}{\metre}. They violate the linear trajectory assumption implicit in many solvers. In extreme cases, Earth grazing fireballs may last for tens of seconds and their path (ignoring deceleration) is a hyperbola with respect to the Earth's surface, requiring a special approach to solve \citep[e.g. ][]{ceplecha1979earth, sansom20193d}. Here we explore which trajectory solver is best for these long fireballs and investigate the influence of gravitational bending on the radiant precision.

As the parameter space of possible radiants and lengths of possible meteorite-dropping fireballs is beyond the scope of this paper, we use as a single case-study a specific  fireball that was observed above Southwestern Ontario on September 23, 2017 by 3 stations of the SOMN. The fireball was first observed at \SI{75}{\kilo \metre} and ended at \SI{35}{\kilo \metre}, lasting \SI{6.5}{\second}. It entered the atmosphere at an angle of \ang{30} from the ground and with an initial velocity of \SI{14}{\kilo \metre \per \second}.

Simulations were performed using the estimated radiant and the velocity, but the initial positions of the fireball were randomly generated inside the fields of view of the simulated all-sky network to cover a wide range of geometries. The dynamics of the fireball were simulated by using the linear deceleration meteor propagation model. The deceleration prior to detection was not included in the simulation because these bright fireballs do not significantly decelerate prior to generating a visible trace in a video system \citep{vida2018modeling}. The duration of the fireball was set to \SI{6.5}{\second}, the time of the beginning of deceleration $t_0$ was randomly generated in the range [0.3, 0.6] of the total duration of the fireball, and the deceleration was randomly generated in the range [1500, 2750] \SI{}{\metre \per \square \second}, which was comparable to the observed event, although the real deceleration was of course not constant.

Due to the large range of simulated starting points, not all of the 100 simulated fireballs were observed in full. Only those simulations in which all stations observed the fireball for at least \SI{4}{\second} where chosen, bringing the number down to 74 simulated meteors. Table \ref{tab:fireballs_solvers_performance} gives the comparison of the performance of various solvers applied to this simulated data. The estimated trajectories were quite precise because of the long duration, thus it was decided to constrain the failure window of interest to $\Delta_{Rmax} = \ang{0.5}$, $\Delta_{Vmax} = \SI{0.5}{\kilo \metre \per \second}$. We justify the reduction of these constraints compared to the SOMN constraints used above by the higher number of observed points (thus better fits), and the fact that \cite{granvik2018identification} indicate that the precision of a fireball's initial velocity should be estimated to around \SI{0.1}{\kilo \metre \per \second} for the statistical distribution of initial source regions to be stable.

\begin{table}
	\caption{Trajectory solver performance for a simulated long-duration fireball observed by a three station all-sky system. The trajectory was incorporated in the simulation statistics if the convergence angle was larger than \ang{5} - we use a lower threshold in this case due to the larger average duration of the event, which in turn means that there are more points for trajectory estimation. F is the percentage of failures (total of 74 meteors), i.e. the percentage of radiants that were outside the $\Delta_{Rmax} = \ang{0.5}$, $\Delta_{Vmax} = \SI{0.5}{\kilo \metre \per \second}$ window.}
    {
	\begin{tabular}{l | c | c | c}
	\hline\hline 
	Solver           & \multicolumn{3}{c}{Fireball} \\
	                 & F (\percent) & $\sigma_R$ & $\sigma_V$ \\
	\hline
IP                & 22\percent & \ang{0.18} & \SI{0.03}{\kilo \metre \per \second}\\
LoS               &  8\percent & \ang{0.07} & \SI{0.02}{\kilo \metre \per \second}\\
LoS-FHAV          & 11\percent & \ang{0.13} & \SI{0.02}{\kilo \metre \per \second}\\
Monte Carlo       & 12\percent & \ang{0.09} & \SI{0.03}{\kilo \metre \per \second}\\
MPF const         & 92\percent & \ang{0.32} & \SI{0.07}{\kilo \metre \per \second}\\
MPF const-FHAV    & 89\percent & \ang{0.22} & \SI{0.07}{\kilo \metre \per \second}\\
MPF linear        & 50\percent & \ang{0.09} & \SI{0.02}{\kilo \metre \per \second}\\
MPF exp           & 88\percent & \ang{0.37} & \SI{0.24}{\kilo \metre \per \second}\\
	\hline 
	\end{tabular}
	}
	\label{tab:fireballs_solvers_performance}
\end{table}

\begin{figure}
  \includegraphics[width=\linewidth]{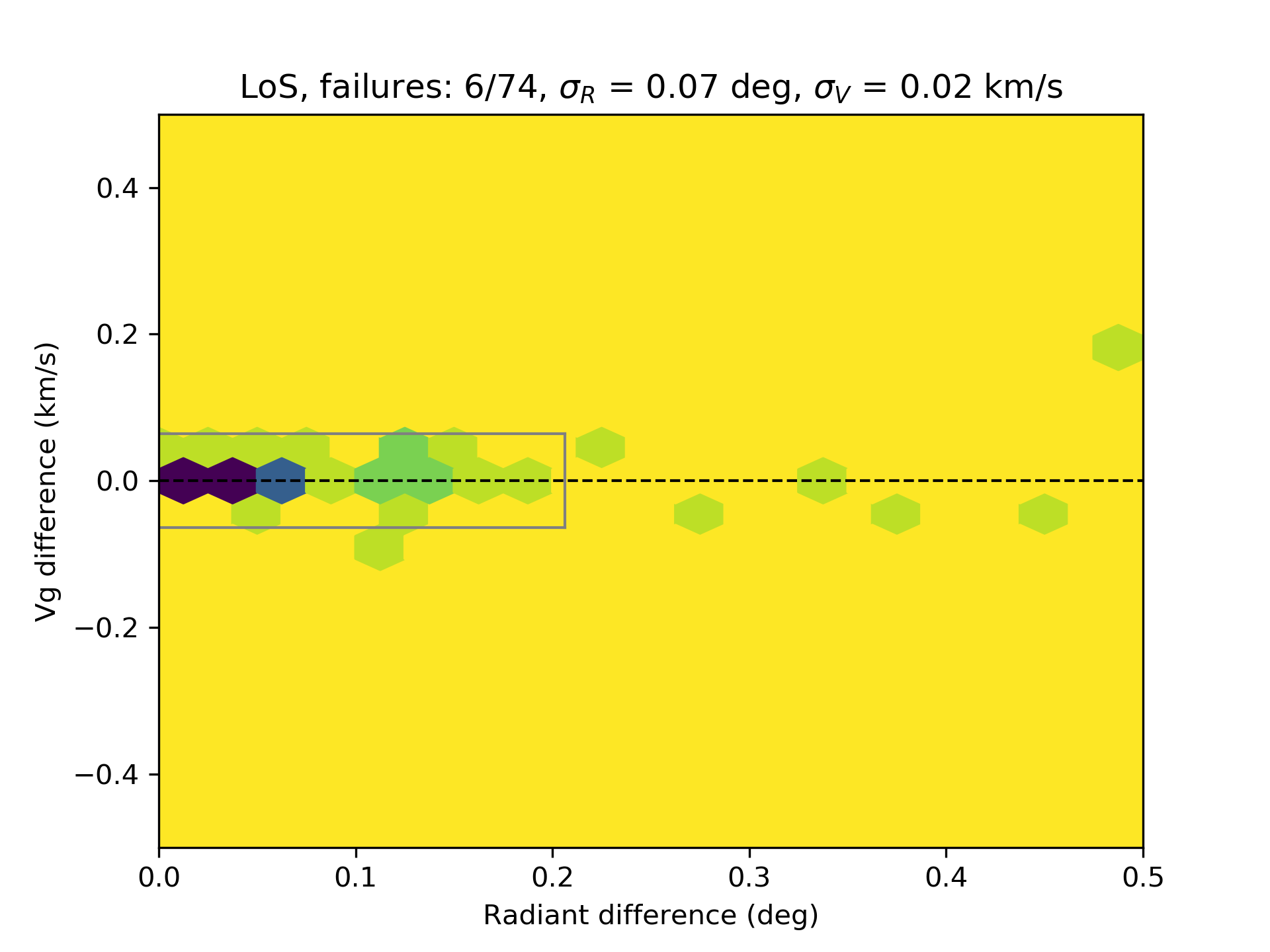}
  \caption{Accuracy of geocentric radiants for simulated fireballs. The solutions were done using the LoS method.}
  \label{fig:long_fireball_los}
\end{figure}

The best performing solver was the \cite{borovicka1990comparison} LoS method with the initial velocity estimated using our newly proposed sliding fit and including compensation for gravity. The expected radiant precision is around 4 arc minutes, while the initial velocity can be estimated to within \SI{20}{\metre \per \second}. Figure \ref{fig:long_fireball_los} shows the 2D histogram of errors for all estimated trajectories. The Monte Carlo solver performs slightly worse, but we haven't manually chosen the solution based on the existence of directionality of the $f_{\Delta t}$ function, as proposed in Paper 1. On the other hand, MPF solvers have a high failure rate, even the MPF solver with the linear deceleration model which should have been able to exactly estimate the trajectory parameters.

\begin{figure}
  \includegraphics[width=\linewidth]{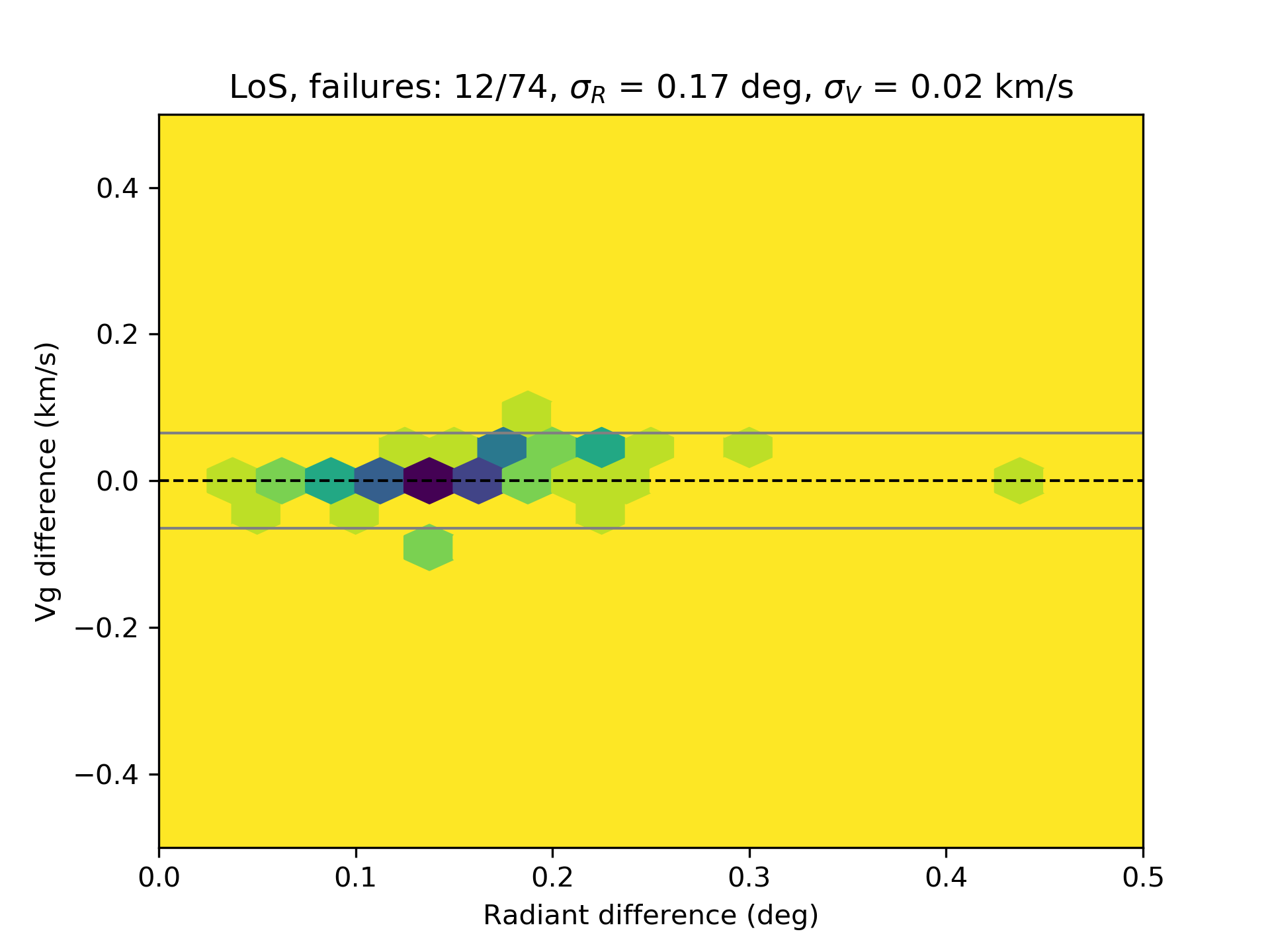}
  \caption{Accuracy of geocentric radiants for a simulated long-duration fireball. The solutions were done using the LoS method, with the option of compensating for the curvature due to gravity turned off.}
  \label{fig:long_fireball_los_nograv}
\end{figure}

Next, we investigated the radiant accuracy if the compensation for the curvature due to  gravity was not taken into account, i.e. the term $\Delta h (t_{kj})$ was kept at 0 (see Paper 1 for details). Figure \ref{fig:long_fireball_los_nograv} shows an offset of about \ang{0.15} from the true radiant, caused by the shift of the estimated radiant towards the local zenith. We point out that these results were obtained on simulated data which ignore any other forces acting on the meteoroid (which are expected to be negligible in any case), which demonstrate that the curvature of the trajectory due to gravity should be compensated for directly during trajectory estimation, as it otherwise produces a significant bias in the direction of the radiant. This becomes more pronounced with the increasing duration of the fireball. Alternatively, an analytic zenith attraction correction could be developed (separate from the zenith attraction correction for computing geocentric radiants) which is dependent on the geometry and the duration of the fireball.

\subsection{Estimated radiant error and true accuracy} \label{subsec:error_analysis}

In this final section we investigate if the radiant error estimated as the standard deviation from the mean accurately describes the true magnitude of the error. We do not investigate the velocity error estimate, because the accuracy of pre-atmosphere velocity for smaller meteoroids is entirely driven by deceleration prior to detection \citep{vida2018modeling}, causing a systematic underestimate of top of atmosphere speeds.

Using the results of the Monte Carlo solver, we compute the angular separation $\theta$ between the estimated and the true geocentric radiant used as simulation input, and divide it by the hypotenuse of standard deviations in right ascension and declination:

\begin{equation}
    \sigma_{err} = \frac{\theta}{\sqrt{(\sigma_{\alpha} \cos \delta)^2 + \sigma_{\delta}^2}}
\end{equation}{}

\noindent The value of $\sigma_{err}$ indicates how many standard deviations from the mean the true radiant is. In an ideal case where the errors would be correctly estimated, the distribution of $\sigma_{err}$ values for all trajectories would follow a truncated normal distribution with a standard deviation of one.

\begin{figure}
  \includegraphics[width=\linewidth]{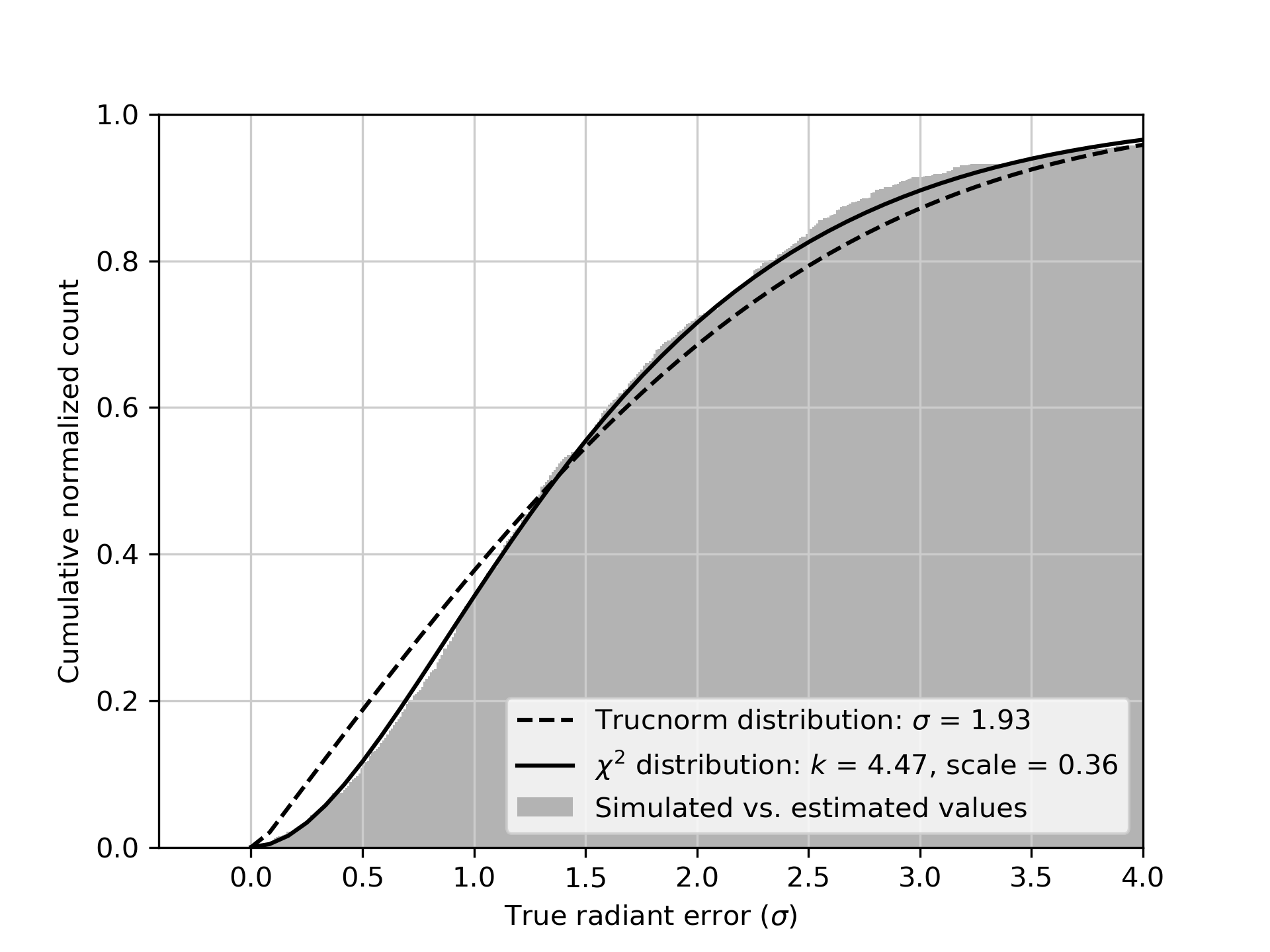}
  \caption{Quality of error estimation for a CAMS-like system and 1000 Geminids.}
  \label{fig:cams_gem_true_vs_estimated_error}
\end{figure}

In practice, the errors seem to be on average underestimated by a factor two across all nine combinations of systems and showers. Figure \ref{fig:cams_gem_true_vs_estimated_error} shows the cumulative histogram of $\sigma_{err}$ values for 1000 simulated Geminids for a CAMS-like system. We fit a truncated normal distribution using the maximum-likelihood method to $\sigma_{err}$ values and report standard deviations for all showers and systems in table \ref{tab:estimated_vs_true_errors}. It appears that the truncated normal distribution is not representative of the underlying distribution of $\sigma_{err}$ values, although its standard deviation is a rough proxy for the scale of the error underestimation. We also fit a $\chi^2$ distribution which appears to be better suited (p-values are consistently high).

\begin{figure}
  \includegraphics[width=\linewidth]{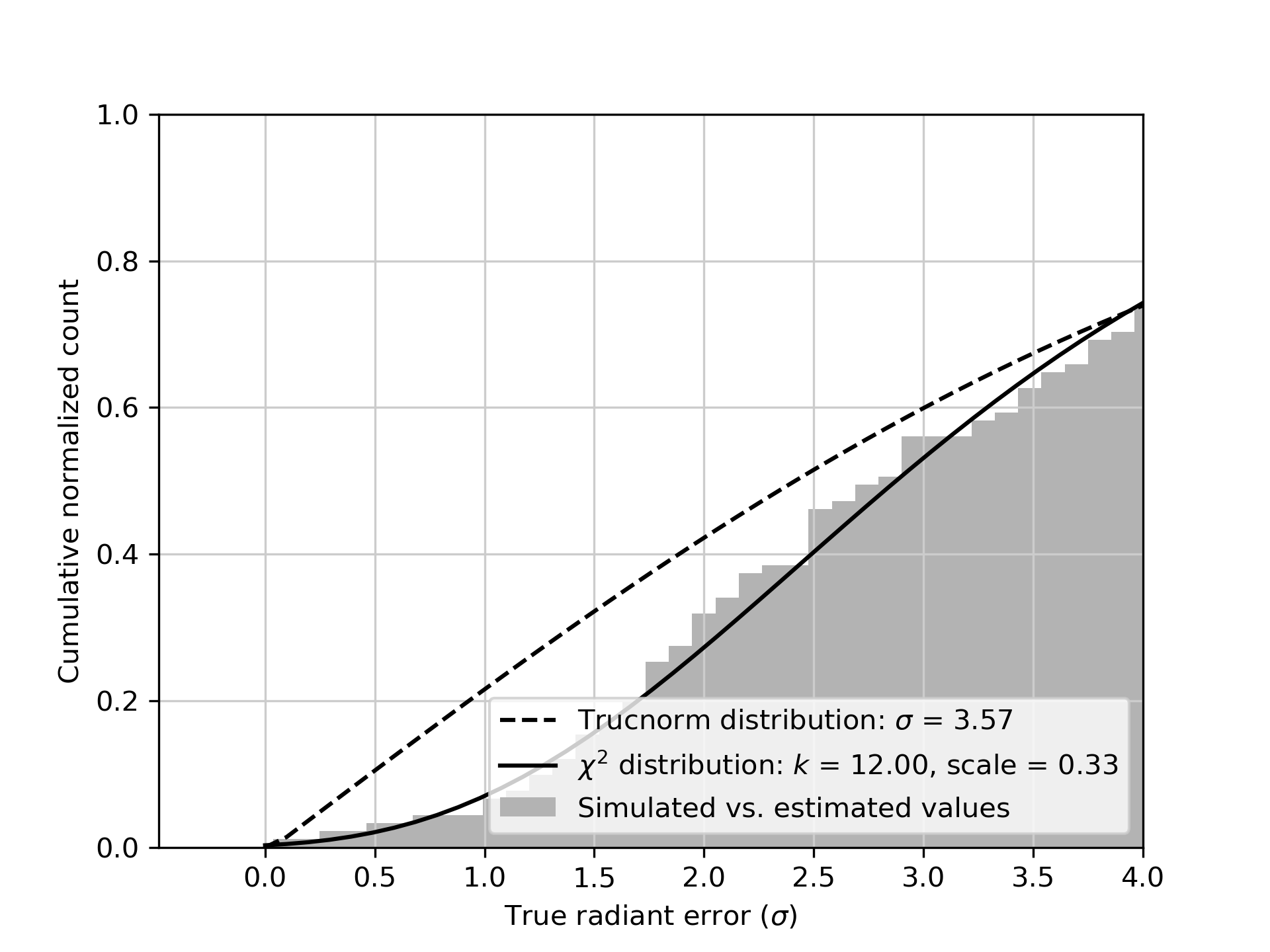}
  \caption{Quality of error estimation for CAMO and 100 Draconids.}
  \label{fig:camo_dra_true_vs_estimated_error}
\end{figure}

Regarding CAMO, the real radiant errors are 2 to 4 times larger than estimated, as seen in figure \ref{fig:camo_dra_true_vs_estimated_error} which shows the error analysis for simulated Draconids. These results indicate that for a robust understanding of errors, a detailed analysis must be done for each system and shower using the shower simulator.

\begin{table*}
	\caption{Parameters of fitted truncated normal and $\chi^2$ distributions to values of $\sigma_{err}$ for different simulated showers and systems. $\sigma$ is the standard deviation of fitted truncated normal distributions (it can be considered as a rough proxy for the magnitude of error underestimation), and "k" and "scale" are parameters of the $\chi^2$ distribution.}
	{
	\begin{tabular}{l | c | c | c | c | c | c}
	\hline\hline 
	Shower	& \multicolumn{2}{c|}{DRA} 					& \multicolumn{2}{|c|}{GEM}					& \multicolumn{2}{|c}{PER} 					\\
			& $\sigma$ 	& k, scale & $\sigma$  & k, scale & $\sigma$ 	& k, scale \\
	\hline
	All-sky	& 2.37					&  3.23, 0.57		& 2.03					& 3.99, 0.42		& 2.43					& 4.70, 0.45 		\\
	CAMS	& 2.02					&  4.61, 0.37		& 1.93					& 4.47, 0.36		& 2.02					& 6.37, 0.28 		\\
	CAMO	& 3.57					& 12.00, 0.33		& 3.74					& 4.82, 0.67		& 2.35					& 2.84, 0.61 		\\
	\hline 
	\end{tabular}
	}
	\label{tab:estimated_vs_true_errors}
\end{table*}

\section{Conclusion}

In Paper 1 \citep{vida2019meteortheory} we described and implemented in Python several trajectory solvers. In this paper we have applied each solver in turn to synthetic meteors with radiants, speeds, and physical properties appropriate to the Draconids, Geminids and the Perseids as they would be recorded by various simulated meteor observation systems. While we have generically investigated solver performance for common optical systems, the simulator allows for detailed simulation of individual real-world meteor observing systems and estimation of the measurement accuracy of meteor showers or individual events of interest. While we summarize some major trends of our simulation comparisons, it is important to emphasize that these results pertain only to the geometry and number of cameras assumed for each simulated system and shower. Ideally, for real observations, simulations would be repeated for specific geometries and camera systems on a per event basis.

With this caveat in mind, based on our simulations, the following is a summary of which trajectory solver performed best for our chosen showers for each meteor observation system:

\begin{itemize}
    \item All-sky systems (SOMN-like) - As meteor deceleration is not usually seen by these systems, the MPF method with the constant velocity model produces the most robust fits and estimates radiants the most precisely, to within \ang{0.25}. The method significantly underestimates the initial velocity, but if a correct deceleration correction is applied an accuracy of \SI{\sim 250}{\metre \per \second} could be achieved.
    \item Moderate field of view systems (CAMS-like) - The intersecting planes and the lines of sight methods produce good results overall when employed in conjunction with more advanced methods of initial velocity estimation, because meteor deceleration becomes visible at the resolutions of such systems. The Monte Carlo solver results are comparable to these solvers and does not provide further improvement of the solution except for meteors with low convergence angles. We recommend using this solver operationally for these systems. The MPF methods improve the radiant precision, but their estimates of the initial velocity are a factor or 2 worse than with other methods. The expected average radiant and velocity accuracy is around \ang{0.1} and \SI{100}{\metre \per \second}, provided the pre-detection deceleration correction from \cite{vida2018modeling} is used.
    \item The CAMO, high-precision system which observes meteor dynamics (deceleration) well - The Monte Carlo and LoS solvers perform the best. The expected average radiant accuracy is around \ang{0.01} and \SI{50}{\metre \per \second}, provided the pre-detection deceleration correction from \cite{vida2018modeling} is used. The MPF approach enforces meteor propagation models which are mismatched to the actual deceleration behavior, resulting in fits with larger errors for these high angular and temporal resolution measurements. 
    \item Meteoroid physical properties strongly influence the velocity accuracy. Meteoroids of asteroidal origin have a factor of 2 higher velocity uncertainties due to larger deceleration, a conclusion previously reported in \cite{vida2018modeling}.
\end{itemize}

We show that a minimum radiant accuracy of order 3 to 6 arc minutes (\ang{0.05} - \ang{0.1}) is needed to measure the true radiant dispersion of younger meteor showers. This value was derived by simulating the 2011 Draconids outburst, a year where the encounter with recently ejected meteoroids having a very low dispersion. With the use of an appropriate trajectory solver, this accuracy can be achieved using moderate FOV and CAMO resolution systems, i.e. systems with the angular resolution better than 3 arc-minutes per pixel (assuming a real precision of around 1 arc-minute is achievable through centroiding).
 
Simulation of the accuracy of low-precision all-sky systems with approximately 20 arc-minute per pixel angular resolution, shows that they are not precise enough to observe structures in meteor showers such as the Taurid resonant branch \citep{spurny2017discovery}. For accurate orbital measurements we strongly suggest installation of more precise all-sky systems with angular resolutions approaching or exceeding one arc-minute per pixel so that the velocity accuracy of less than \SI{0.1}{\kilo \metre \per \second} can be achieved, as recommended by \cite{granvik2018identification}. We show that compensation for trajectory bending due to gravity should be taken into account for longer fireballs (> 4 seconds) due to its significant influence on the radiant accuracy, as noted earlier by \cite{ceplecha1979earth} and \cite{sansom20193d}.

Finally, we investigated the quality of our radiant error estimation approach by comparing estimated errors to known absolute error from the simulation input. We find that radiant errors are underestimated by a factor of 2 for all-sky and CAMS-like systems, and by a factor of 3 to 4 for CAMO.

\subsection{Note on code availability}

Implementation of the meteor simulator as well as implementation of all meteor solvers used in this work are published as open source on the following GitHub web page: \url{https://github.com/wmpg/WesternMeteorPyLib}. Readers are encouraged to contact the authors in the event they are not able to obtain the code on-line.

\section{Acknowledgements}
We thank Dr. Eleanor Sansom for a helpful and detailed review of an earlier version of this manuscript. Also, we thank Dr. Auriane Egal for suggestions about the modelling of the Draconids. Funding for this work was provided by the NASA Meteoroid Environment Office under cooperative agreement 80NSSC18M0046. PGB acknowledges funding support from the Canada Research Chair program and the Natural Sciences and Engineering Research Council of Canada.




\bibliographystyle{mnras}
\bibliography{bibliography} 


\section{Simulation orbit reports}

In supplementary files we provide all trajectory and orbit report files of simulated trajectories used in this paper. The report files include all inputs to the solver, as well as all trajectory and orbit parameters. We encourage readers to use these values to verify their implementations of the method.


\bsp	
\label{lastpage}
\end{document}